\documentclass[sigconf,edbt]{acmart-edbt2024}
\usepackage{listings}
\usepackage{capt-of}
\usepackage[ruled,linesnumbered,noend]{algorithm2e} 

\newcommand{\URI}[0]{\mbox{$\mathcal{U}$}}
\usepackage{array}
\newcommand{\T}[0]{\mbox{$\mathcal{T}$}}

\newcommand{\Blank}[0]{\mbox{$\mathcal{B}\mathcal{N}$}}
\newcommand{\Lit}[0]{\mbox{$\mathcal{L}$}}
\usepackage{algpseudocode}
\usepackage[utf8]{inputenc}
\usepackage{color}
\usepackage{hyperref}
\usepackage{graphicx}
\usepackage{placeins}
\usepackage{float}
\usepackage{subfigure}
\usepackage{fancyvrb}
\usepackage{xcolor}
\usepackage{xspace}
\newcommand{\CG}[0]{ChatGPT\xspace}

\newcommand{\new}[1]{{\color{black}#1}}
\newcommand{\neo}[1]{{\color{black}#1}}
\makeatletter
\newcommand{\removelatexerror}{\let\@latex@error\@gobble}
\makeatother
\usepackage{balance}

\usepackage{graphicx}

\usepackage{hyperref}

\def\BibTeX{{\rm B\kern-.05em{\sc i\kern-.025em b}\kern-.08em
    T\kern-.1667em\lower.7ex\hbox{E}\kern-.125emX}}

\usepackage{booktabs} 

\setcopyright{rightsretained}

\acmDOI{}

\acmISBN{XXXXXXXXXXXX}

\begin{document}






\title{
Validating ChatGPT Facts \neo{through} \\RDF Knowledge Graphs and Sentence Similarity}

\author{Michalis Mountantonakis and Yannis Tzitzikas}
\email{{mountant|tzitzik}@ics.forth.gr}
\affiliation{
  \institution{FORTH-ICS and Computer Science Department, University of Crete}
  \country{Greece}
}


\begin{abstract} 
Since \CG offers detailed responses without justifications, and erroneous facts even for popular persons, events and places, in this paper we present a novel pipeline that retrieves the response of \CG in RDF and tries to validate the \CG facts using one or more RDF Knowledge Graphs (KGs). To this end we \neo{leverage} DBpedia and LODsyndesis 
\neo{(an aggregated Knowledge Graph that contains 2 billion triples from 400 RDF KGs of many domains)}  
and short sentence embeddings, 
and \neo{introduce} an algorithm that returns the more relevant triple(s) accompanied by their provenance and a confidence score.
\neo{This  enables the validation of ChatGPT responses and their enrichment with  justifications and provenance.}
To evaluate \neo{this service (such services in general)}, we create an evaluation benchmark that includes \new{2,000 \CG facts; 
specifically 1,000 facts for famous Greek Persons, 500 facts for popular Greek Places, and 500 facts for Events related to Greece. The facts were manually labelled (approximately 73\% of \CG facts were correct and 27\% of facts were erroneous).}
The results are promising; indicatively for the whole benchmark, we managed to verify the 85.3\% of the correct facts of \CG and to find the correct answer for the 58.0\% of the erroneous \CG facts.
\end{abstract}






\maketitle
\section{Introduction}

\CG\ is a novel Artificial Intelligence (AI) chatbox \cite{OpenAI}, which is built on GPT-3.5 and GPT-4 families of large language models (LLMs) \cite{brown2020language}, and provides detailed responses and human-like answers across many domains of knowledge. 
However,  as it is also stated in its webpage, ``ChatGPT may produce inaccurate information about people, places, or facts" \cite{van2023chatgpt},
\neo{
and it has ``limited knowledge of world and events after 2021". }

\begin{figure*}[h!]
    \centering
\fbox{\includegraphics[width=0.75\linewidth]{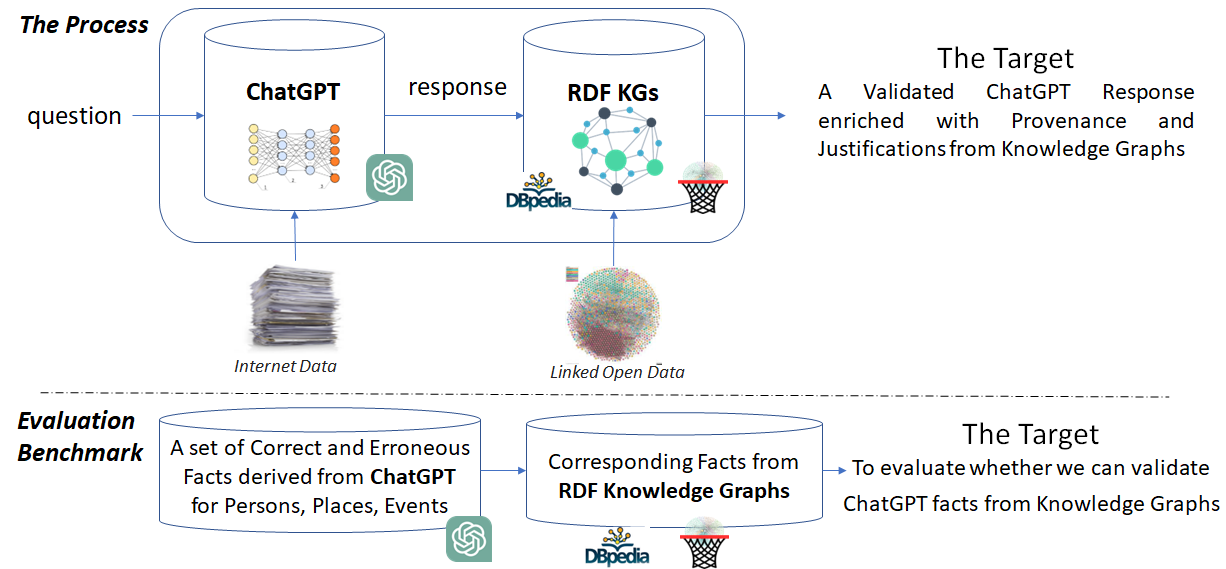}}
\caption{The key notion of Combining \CG with RDF KGs.}
    \label{fig:objective}
\end{figure*}

\neo{
One challenge, as \new{shown in Fig. \ref{fig:objective}, is} how to combine \CG\  \new{which has been trained by using data from web sources (such as Wikipedia, books and news articles),} 
with popular RDF KGs 
(Knowledge Graphs \cite{hogan2021knowledge} expressed in RDF\footnote{\url{https://www.w3.org/RDF/}}). The objective is to
enable the validation of \CG responses 
and their enrichment with justifications and provenance information, 
given the high quality of information of RDF KGs 
and that most of them are updated either periodically or continuously \cite{farber2018linked}.  Another challenge is how to enable the evaluation of the mentioned process, given that there are no available such benchmarks, i.e., benchmarks containing \CG facts that can be validated through RDF KGs  (lower side of   Fig. \ref{fig:objective}).
}

\begin{figure*}[t]
    \centering
\fbox{\includegraphics[width=0.78\linewidth]{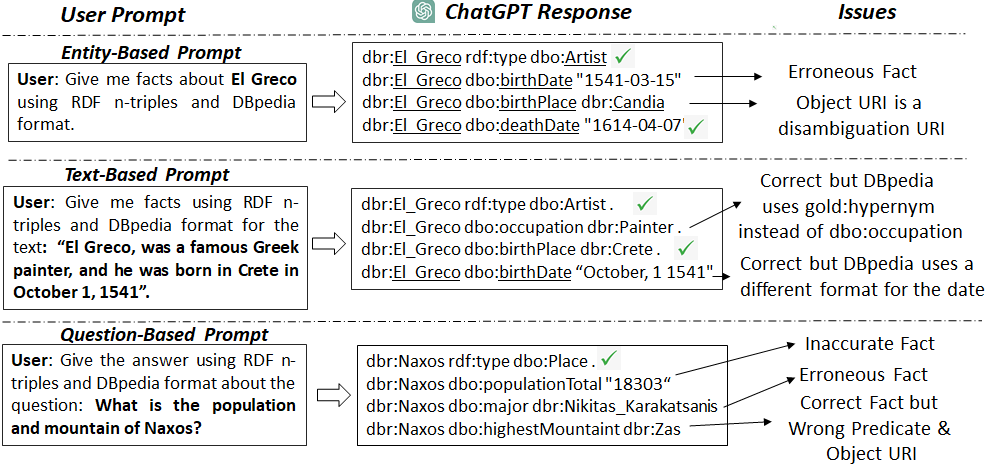}}
\caption{\CG responses in RDF format using DBpedia Model and Common Errors}
    \label{fig:context}
\end{figure*}

\neo{
ChatGPT can be used for producing RDF triples, e.g. see the examples of Fig. \ref{fig:context},
however, in many cases
it produces 
erroneous facts and URIs \cite{mountantonakis2023using}, without giving any evidence about the provenance of 
provided information.
}
%
%
In particular, there are several ways to ask about RDF N-triples \cite{beckett2014rdf}.
Although in many cases the triples are exactly the same as in DBpedia 
there are several \neo{problematic cases}  that can occur and some of them are described below. 
First, the triple can be valid \neo{syntactically} but the 
\neo{denoted}
fact can be erroneous or inaccurate, e.g., see the case of the birth date of the painter ``El Greco" or the 
\neo{mountain} of ``Naxos" island in Fig. \ref{fig:context}. Second, the fact can be correct and the URIs valid, however, \CG can fail to produce the correct URIs for the subject, predicate or object, e.g., see that in some cases the Object URI is a disambiguation page. Third,  the fact can be correct, but one or more URIs can be invalid or not the appropriate one, e.g., in Fig. \ref{fig:context},  the predicates dbo:occupation and dbo:highestMountain are returned from \CG, however for the corresponding triples in DBpedia, the predicates gold:hypernym and dbp:highestMount are used. 
Finally, 
\neo{there are cases where}
the literals are not exactly the same, different formats are used for the literals (e.g., see the birth date in the text-based prompt),  and literals are used instead of URIs (or the opposite), e.g., for the last fact of Fig. \ref{fig:context}, the corresponding triple in DBpedia contains the literal ``Mt. Zeus" instead of dbr:Zas. 

\neo{
To combine \CG with KGs, 
}
we need to tackle the aforementioned \CG limitations.
\neo{This raises the following research questions (RQs):}  

$\bullet$ \textbf{RQ1}: How to check the validity of \CG facts 
by using RDF KGs
for validating the correct \CG facts and for finding the correct answer for its erroneous facts? 

$\bullet$ \textbf{RQ2}: Is it effective to use \neo{several}  KGs (e.g., through LODsyndesis \cite{mountantonakis2021services}) for checking the validity of facts instead of using only DBpedia, 
\new{given that more facts can be validated through the complementary information, however more conflicts can occur (e.g., in case of contradicting and outdated data from KGs)}?

As regards our contribution, we propose a novel pipeline that receives as input a \CG response in RDF format and tries to find the most similar fact in a KG for enabling the validation of any \CG fact. For performing this process, we use two different KGs, i.e., DBpedia \cite{lehmann2015dbpedia} and LODsyndesis \cite{mountantonakis2021services} (it contains 400 RDF KGs including DBpedia). The proposed algorithm sends SPARQL queries and REST requests for retrieving candidate similar triples and creates short sentence  embeddings for finding the most similar triple to each \CG fact according to the cosine similarity score of vectors. 
Concerning the evaluation, we have created a benchmark containing 2,000 facts from \CG,  for a list of i) famous Greek Persons (from the Ancient and Modern era),  ii) popular Greek Places, including Cities, Islands, Lakes, Mountains and Heritage Sites, \new{and iii) Events related to Greece, such as Battles, Sport Events, Earthquakes and Elections.} Moreover, we have manually labelled the \CG facts as correct or erroneous (approximately 26\% of \CG facts were erroneous) and we provide an experimental evaluation by comparing the effectiveness and efficiency of using one or more KGs for validation.
As regards the results, indicatively, by using LODsyndesis we managed to verify $92.2\%$ of the correct \CG facts for Persons, $77.1\%$ for Places, \new{and  $76.3\%$ for Events,} whereas we found the correct answer for the $57.4\%$ of the  erroneous \CG facts for Persons, $49.1\%$ for Places \new{and $68.2\%$ for Events.}

\begin{figure*}[t]
    \centering
\fbox{\includegraphics[width=0.93\linewidth]{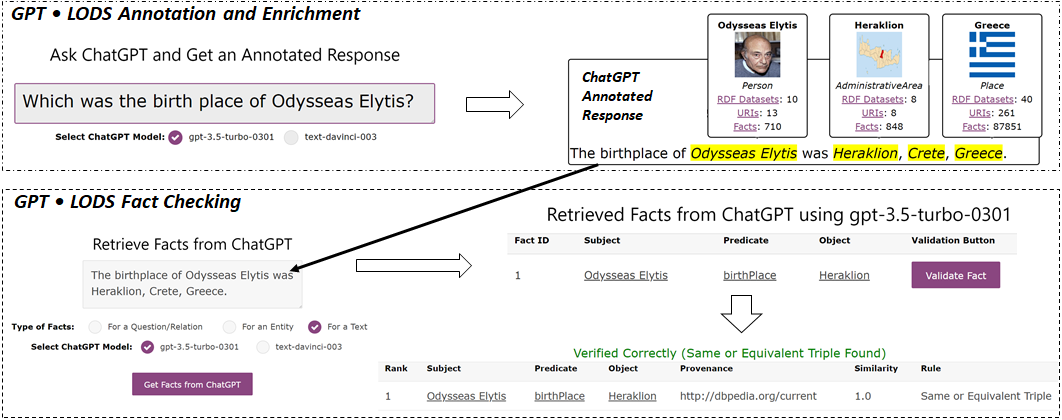}}
\caption{A scenario showing a use case that exploits the presented Fact Checking approach}
    \label{fig:useCase}
\end{figure*}

Moreover, we have created a demo web application with name GPToLODS, which is available online\footnote{\url{https://demos.isl.ics.forth.gr/GPToLODS/}}. Through this website, one can have access to both a fact checking service \cite{mountantonakis2023iswc} that exploits the pipeline of this paper,  and also to an annotation and enrichment service \cite{mountantonakis2023using}. 
Concerning the use cases that are related to the mentioned approach \cite{mountantonakis2023iswc},  the most important are the following: (a) QA and Fact Validation, and (b) Relation Extraction and Triples Generation, even for texts including facts that are not part of a KG. Indicatively, for the first use case, we present in Fig. \ref{fig:useCase} a scenario including all the services of GPToLODS, whereas all the use cases are presented in an online video\footnote{\url{https://www.youtube.com/watch?v=5DW1d37aPMc}}.

In particular, the user (who is not required to be familiar with RDF) asks a question in natural text (see the example in Fig. \ref{fig:useCase}), e.g., ``Which was the birth place of Odysseas Elytis?" (a Nobel Prize-winning Greek poet), and the annotation and enrichment service of GPToLODS, returns the \CG response with annotations, images and links to RDF KGs (upper side of  Fig. \ref{fig:useCase}). The user can validate the information by using the fact checking service. In particular, the returned textual \CG response is converted to triples by sending to \CG the following prompt: ``Give me facts using RDF N-triples and DBpedia format for the text: The birthplace of Odysseas Elytis was Heraklion, Crete, Greece" (see the lower left part of Fig. \ref{fig:useCase}). Afterwards, the facts are retrieved from \CG, and they are validated by using the pipeline of this paper and the selected KG(s). Finally, the user can see the corresponding facts from the KG(s) and their provenance (in the lower right side of Fig. \ref{fig:useCase} we  confirmed the fact from DBpedia).

Concerning the novelty, to the best of our knowledge, this is the first work trying to validate the facts of \CG by using one or more RDF KGs, whereas it also offers an evaluation benchmark for fact checking \new{over \CG with 2,000 facts}.

The rest of this paper is organized as follows: \S \ref{sec:backrw} discusses the background and the related work, \S \ref{sec:process} introduces the proposed process by showing all the steps and the algorithm for finding relevant facts. Moreover, \S \ref{sec:benchmark} presents the evaluation benchmark and statistics, whereas \S \ref{sec:eval}  presents the experimental evaluation over the benchmark. Finally, \S \ref{sec:conc} concludes the paper and discusses future directions.

\section{Background \& Related Work}
\label{sec:backrw}
This section discusses the background (see \S \ref{sec:back}) and the related work (see \S \ref{sec:rw}).

\subsection{Background}
\label{sec:back}

\neo{
An RDF triple,
is  a triple the form subject-predicate-object $\langle$s,p,o$\rangle$
that belongs to the set of all possible triples  $\T$,
defined as 
$\T$=$(\URI \cup \Blank) \times (\URI) \times (\URI \cup \Blank \cup \Lit)$, where $\URI$, $\Blank$ and $\Lit$ denote the sets of URIs, blank nodes and literals, respectively.
A KG (e.g., DBpedia or LODsyndesis) is essentially  a subset $T_{KG}$ of $\T$.
}


We shall use $[x]$ to denote the  class of equivalence of an element $x$ ($x \in \URI$ or $x \in \Lit$). Specifically, two or more  URIs (resources, properties or RDF classes) belong in the same class of equivalence if they refer to  the same real world entity (as inferred by the owl:sameAs), property or class, whereas two or more  literals are equal, if they use exactly the same string. 
Moreover, we define that two triples $t=\langle s,p,o\rangle,t^{\prime}=\langle s^{\prime},p^{\prime},o^{\prime}\rangle$ are equivalent when it holds that $[s] \equiv[s^{\prime}],~[p] \equiv [p^{\prime}], [o] \equiv[o^{\prime}]$.
 An example of equivalent triples between two KGs, say DBpedia and Wikidata follows: $\langle$dbo:Aristotle,dbo:birthDate,``384 BC"$\rangle$ and  $\langle$wkd:Q868,wkp:P569,``384 BC"$\rangle$, since it holds that  $\langle$dbp:Aristotle,owl:sameAs, wkd:Q868$\rangle$, $\langle$dbo:birthDate, owl:equiv- alentProperty, wkp:P569$\rangle$ and they use the same literal as object, i.e., ``384 BC". Finally, we define all the triples of an entity $e$ in a $KG$ as follows: $T_{KG}(e)=\{\langle s,p,o\rangle \in T_{KG}~|~[s]\equiv e ~or~[o]\equiv e\}$.

\subsubsection{DBpedia and LODsyndesis KG.} In this paper, we  use both DBpedia \cite{lehmann2015dbpedia} and LODsyndesis KG \cite{mountantonakis2021services,mountantonakis2020content} for enabling the validation of \CG facts. LODsyndesis is an Aggregated Knowledge Graph that contains 2 billion triples from 400
RDF KGs of many domains (including cross-domain KGs like DBpedia,  Wikidata \cite{vrandevcic2014wikidata}, YAGO \cite{rebele2016yago}, geographical KGs like GeoNames \cite{geonames}, etc.). In particular, LODsyndesis has computed  the transitive and
symmetric closure of 45 million equivalence relationships, including owl:sameAs, owl:equivalentProperty and owl:equivalentClass relationships, and it has indexed the contents of the 400 RDF KGs, by preserving their provenance.
Due to the precomputed closure, it enables the retrieval of equivalent triples to a given triple, which is quite important given that for a fact of \CG, it is possible to find in the KG a semantically equivalent triple and not the exact one.

\subsection{Related Work}
\label{sec:rw}
Here, we focus on approaches that combine \CG and KGs, fact checking using RDF KGs and embeddings over RDF triples. Finally, we provide a comparison and we mention the novelty of the proposed approach.

\subsubsection{A. \CG and Knowledge Graphs (KGs).} 
First, \cite{omar2023chatgpt} provides a comparison for the Question Answering task, showing that \CG\ can have high precision in popular domains but very low precision in unseen domains.  In \cite{gonzalez2023yes}, the authors evaluated \CG in a named identification task over historical documents, whereas in \cite{hoes2023using} \CG  facts about politics were  manually validated. Moreover, \cite{chatgpt2023rdf} mentions the importance of providing solutions that combine \CG and RDF KGs, e.g., for improving its accuracy. \new{ In \cite{tan2023evaluation}, the authors compared the performance of \CG and traditional models based on Knowledge Bases on 8 complex Question Answering datasets. Regarding the key results, \CG achieved better results compared to the best traditional models for old datasets, however, it was worse for the  current state-of-the-art datasets. In addition,  \cite{yang2023chatgpt} reviews  studies
on enhancing PLMs with KGs, and  proposes to enhance LLMs with KGs by developing
KG-enhanced large language models (KGLLMs), however,  an evaluation has not been yet conducted.

Furthermore, the author in \cite{summarygpt} exploited \CG for summarizing RDF KGs by using a \CG constructed classifier, whereas  \cite{shi2023chatgraph}
proposed an interpretable text classification
framework, which can extract refined and structural knowledge from natural text through a KG extraction task, by utilizing \CG. In \cite{trajanoska2023enhancing}, a Natural Language Processing method that exploits \CG was presented  for enabling the construction of a KG on the
topic of sustainability using raw documents. Moreover, in \cite{luo2023chatrule} the authors proposed a Large Language Model based rule generator for mining logical rules over KGs.  In \cite{yao2023exploring} a framework  which can be used for KG completion by exploiting \CG is presented, by considering triples in KGs as text sequences.

Concerning our past work, \cite{mountantonakis2023using} presents an approach where the response of \CG is annotated by using popular Entity Recognition tools, and each entity is enriched with more statistics and links to LODsyndesis, whereas \cite{mountantonakis2023iswc} presents a demo application that exploits the fact checking approach that is described in this paper, for enabling the real validation of \CG facts.}

\subsubsection{B. Fact Checking by using KGs.} There is a high trend for automated fact checking approaches by using both textual sources and KGs \cite{zeng2021automated}. Concerning approaches over KGs, \cite{dong2015knowledge} used Freebase as a gold standard, for identifying the true values for facts extracted from web pages. Also, DBpedia KG was used in \cite{ciampaglia2015computational} for performing fact checking using simple network models and by using input both natural text (from Wikipedia) and structured data. Furthermore, in \cite{shiralkar2017finding}, DBpedia was also used in an unsupervised network-flow based approach for validating statements, whereas \cite{vedula2021face} provides a pipeline for explainable fact
checking through structured and unstructured data. Moreover, \cite{maliaroudakis2021claimlinker} offers a Web service and API that links natural text to a KG of fact-checked claims whereas  \cite{huynh2019buckle} presents a benchmark for comparing and evaluating fact checking algorithms over KGs.

\subsubsection{C. Embeddings using RDF Triples (focusing on Fact Checking).} There are many approaches that create embeddings from RDF Triples (or/and paths) for several tasks \cite{portisch2022knowledge}, including finding similar entities \cite{ristoski2019rdf2vec,mountantonakis2019knowledge} and link prediction \cite{portisch2022knowledge,biswas2020embedding}. Concerning fact checking approaches over RDF, \cite{da2021using}  proposed a path-based
approach that creates KG embeddings, whereas \cite{qudus2022hybridfc}  introduces a hybrid approach which uses both text and KG embeddings by using DBpedia. Also, in \cite{pister2019knowledge,ammar2019fact} embeddings from RDF triples were used for  fact validation for a Semantic Web Challenge, whereas \cite{huang2022trustworthy} leverages noisy data from web pages and prior knowledge from KGs (using KG embeddings) for checking the veracity of data.
Finally, there are available several evaluation benchmarks for fact checking that are based on popular KGs like DBpedia and Freebase, e.g., \cite{gerber2015defacto,huynh2018towards}.

\subsubsection{Comparison and Novelty.} First, compared to approaches over \CG, we mainly focus on enabling the validation of \CG facts, and not on tasks like
Entity Recognition \cite{mountantonakis2023using} and Question Answering \cite{omar2023chatgpt}, or only on performing a manual fact checking evaluation (e.g., \cite{hoes2023using}). Regarding fact checking, we focus on \CG responses and not on data from textual sources (e.g., \cite{dong2015knowledge,huang2022trustworthy}), whereas we provide an approach that combines both SPARQL queries and the creation of embeddings from multiple RDF datasets (through LODsyndesis KG), instead of using a single one like DBpedia (e.g., \cite{ciampaglia2015computational,shiralkar2017finding}).
Moreover, our benchmark contains data from \CG and not from other types of sources \cite{gerber2015defacto,huynh2018towards}. \new{Finally, compared to our previous work, in \cite{mountantonakis2023iswc} we presented a fact checking service that exploits the approach that is described in this paper, however, in that demo paper we neither  provide details about the algorithm nor an experimental evaluation.}

Regarding the novelty, to the best of our knowledge, this is the first work that a)  validates at real time the facts of \CG by using multiple RDF KGs and short sentence embeddings, and by returning the provenance for each fact, and b) offers a benchmark for fact checking over \CG and RDF KGs, and manually annotated results for facts about real persons, places and events.

\begin{figure*}[t]
    \centering
\fbox{\includegraphics[width=0.65\linewidth]{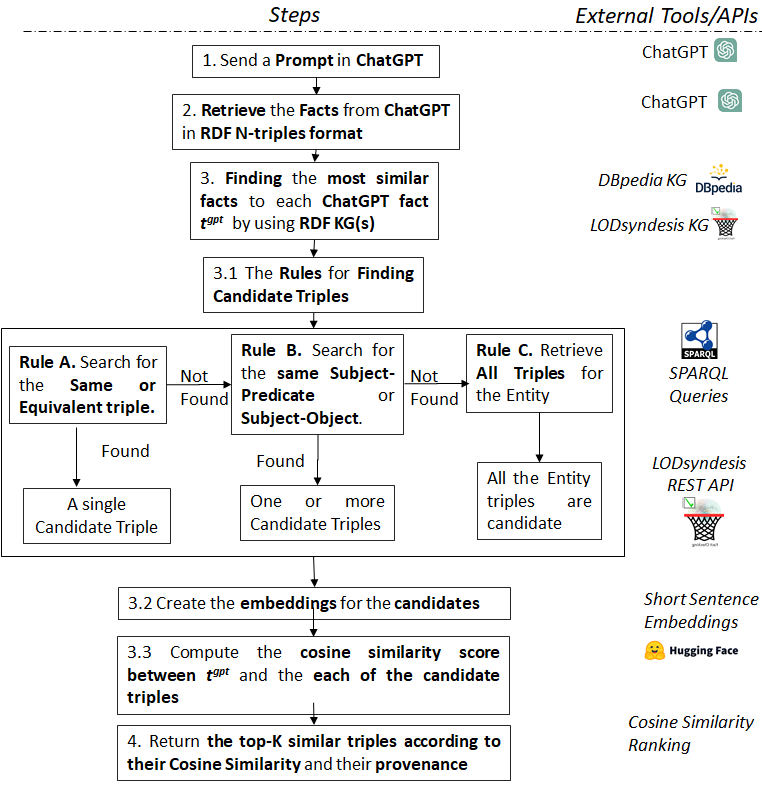}}
\caption{The Steps of the Proposed Approach and external tools/APIs}
    \label{fig:process}
\end{figure*}

\section{The Steps of the Approach}
\label{sec:process}
The steps of the process (depicted in Fig. \ref{fig:process}) are the  following: 1) send a prompt in \CG by asking for RDF triples either from a given response or by asking \CG for facts using the desired format, 2) receive the response as RDF triples, 3) for each fact find the most relevant triple(s) according to three rules by using short sentence embeddings, and  rank them in descending order according to their cosine similarity score with the vector of the desired fact and 4)  produce the $K$ most similar triples (and their provenance) with their score. The running example of Fig. \ref{fig:runningEx} shows a real response from \CG, the candidate triples for each fact and the K=1 most relevant triple. In particular, the first fact is correct and the triple exists in the KG, the second is inaccurate and the KG contains the same entity-predicate pair and finally the third is correct, however, the correct triple in DBpedia uses different URIs for the predicate and the object.
\neo{Below we describe each step in detail.}

\subsection{Step 1. Send a Prompt to \CG }
We list 3 key ways to create RDF N-triples \cite{beckett2014rdf} through  \CG.

$\bullet$ \textbf{Entity-based Prompt.}  One can write this prompt: ``Give me facts about entity $e$ using RDF N-triples and X format", where $e$ is  the name of the entity and X can be the DBpedia format, and it will return the corresponding triples. Such a real case is shown in the upper side of Fig. \ref{fig:context}.

$\bullet$ \textbf{Text-based Prompt.} One can write this  prompt: ``Give me facts using RDF N-triples and X format for the text Y", where Y can be any text, such as the response of \CG.  An example is shown in the middle side of Fig. \ref{fig:context}.

$\bullet$ \textbf{Question-based Prompt.} One can write the following prompt:  ``Give me facts using RDF N-triples and X format about the question $q$", where $q$ can be any question, e.g., see the lower side of Fig. \ref{fig:context}. \newline

\subsection{Step 2. Retrieve the Facts from ChatGPT in RDF format}
The next step is to  collect all the triples from the \CG response, and to analyze each one. In particular, let $T_{GPT}$ be all the triples from a \CG response. For each 
$t_{gpt}=\langle e,p,o\rangle \in T_{GPT}$,
we \neo{try} finding either the same (or an equivalent) triple, a triple with the same subject-predicate or subject-object,  or the most similar triple, to a given KG, as it is described in Alg. \ref{alg:valid}.

\begin{figure*}[t]
    \centering
\fbox{\includegraphics[width=140mm]{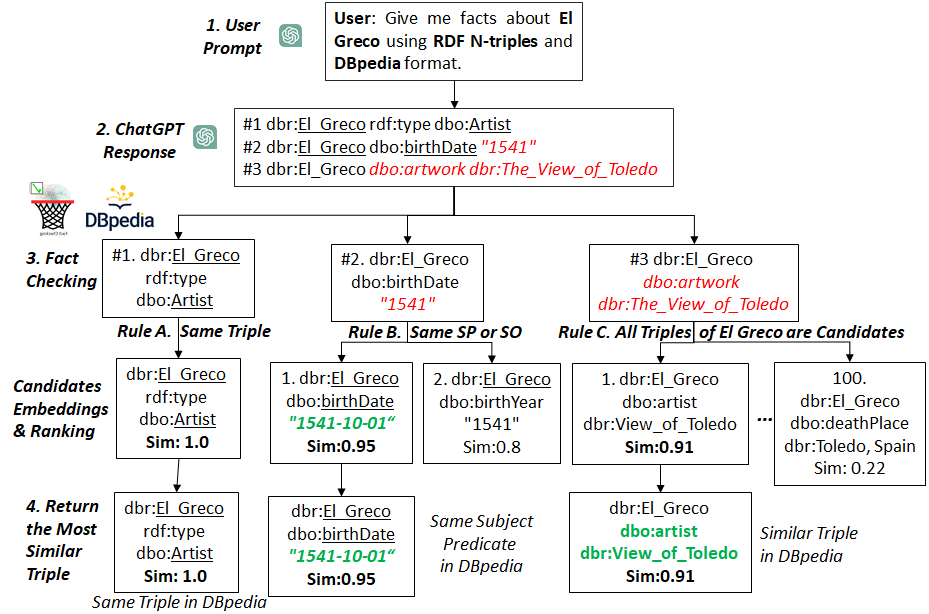}}
\caption{Running Example by using a real response from the \CG}
    \label{fig:runningEx}
\end{figure*}

\removelatexerror
\begin{figure*}[t]
\begin{algorithm}[H]
\SetAlgoNoLine
\Indm
\KwIn{A fact $t^{gpt}=\langle e,p,o\rangle$ from \CG, and a Knowledge Graph $KG$, and a Sentence Similarity model, denoted by $sentenceModel$}
\KwOut{The top-$K$ most similar triples of KG to $t^{gpt}$}
\Indp
$cand(t^{gpt}) \leftarrow \emptyset$ \\ 

\tcp{Execution of Rules for Finding Candidates} 
{\uIf(\tcp*[f]{Rule A.}){$\exists ~ t^{\prime} \in T_{KG}(e)  ~s.t.~ t^{gpt} \equiv t^{\prime} $} { 
         $cand(t^{gpt}) \leftarrow \{t^{\prime}\}$ \tcp*[f]{Same or Equivalent Triple found}  
    }
    \ElseIf(\tcp*[f]{Rule B.}){$\exists~t^{\prime}=\langle e, p^{\prime},o^{\prime}\rangle  \in T_{KG}(e)~ s.t. ~[p^{\prime}]=[p]~or~[o^{\prime}]=[o]$}{
          $cand(t^{gpt})\leftarrow\{t^{\prime}=\langle e, p^{\prime},o^{\prime}\rangle \in T_{KG}(e)~|~[p^{\prime}]=[p]~or~[o^{\prime}]=[o]\}$\tcp*[f]{All the same Subject-Predicate and Subject-Object Pairs are candidates}  
    }
     \Else(\tcp*[f]{Rule C.}){
         $cand(t^{gpt})\leftarrow T_{KG}(e)$\tcp*[f]{All the Triples about $e$ are Candidates}  
     }
}

\tcp{Create the vector for the \CG fact} 
$t^{gpt}_{c} \leftarrow convert(t^{gpt})$ \tcp*[f]{Performing Conversions for the URIs} \\
$\vec{t^{gpt}_{c}}\leftarrow sentenceModel.encode(t^{gpt}_{c})$ \tcp*[f]{Vector for the \CG fact} \\

\ForAll(\tcp*[f]{for each candidate triple}){$t^{\prime} \in cand(t^{gpt})$}
{$t^{\prime}_{c}\leftarrow convert(t^{\prime})$ \tcp*[f]{Performing Conversions for the URIs} \\
$\vec{t^{\prime}_{c}} \leftarrow sentenceModel.encode(t^{\prime}_{c})$ \tcp*[f]{Create the vector for each candidate by using  sentence embeddings}

$\cos(\vec{t^{gpt}_{c}},\vec{t^{\prime}_{c}})\leftarrow{{ \vec{t^{gpt}_{c}}} \cdot  {\vec{t^{\prime}_{c}}}  \over \| {\vec{t^{gpt}_{c}}} \|\| {\vec{t^{\prime}_{c}}} \|}$ \tcp*[f]{Computation of Cosine Similarity}
}
\tcp{Sort the top-$K$ similar triples w.r.t. to the cosine similarity} 
$t_{best}\leftarrow arg_{t^{\prime}}max~\{\cos(\vec{t^{gpt}_{c}},\vec{t^{\prime}_{c}}) ~|~ t^{\prime} \in cand(t^{gpt})\}$ \tcp*[f]{Most Similar fact(s)} \\
\textbf{return} $t_{best}$, $cos(\vec{t^{gpt}_{c}},\vec{t_{best}})$  \tcp*[f]{Return the most similar fact(s) and their similarity score}

\caption{Finding the most similar fact(s) for Validation}
\label{alg:valid}
\end{algorithm}
\vspace{-2mm}
\end{figure*}

\subsection{Step 3.  Algorithm for Finding the most similar facts to each ChatGPT facts by using RDF KG(s)}
The input \neo{of} Alg. \ref{alg:valid} is a \CG fact  $t^{gpt}$,  the KG that will be used, e.g., DBpedia or LODsyndesis, and the embeddings model that will be used. \newline

\subsubsection{Step 3.1. The Rules for Finding Candidate Triples.}
First, we collect all the candidate triples, i.e., $cand(t^{gpt})$ according to three rules (lines 1-7), which are also shown in Fig. \ref{fig:process} and in the running example of Fig. \ref{fig:runningEx}.

\textbf{Rule A. Search for the same or equivalent triple.}
We search in the KGs if the same (or an equivalent) triple of  $t^{gpt}$ exists, i.e., if $\exists~ t^{\prime} \in T_{KG}(e)~s.t.~ t^{\prime} \equiv t^{gpt}$. In this case, $cand(t^{gpt})$ contains the single triple $t^{\prime}$ (lines 2-3).

{\em How the Rule is Executed?} For checking Rule A, we send an ASK SPARQL query to DBpedia SPARQL endpoint \cite{lehmann2015dbpedia} (i.e., when DBpedia KG is used), or a  request to the online Fact Checking Service of LODsyndesis REST API \footnote{\url{https://demos.isl.ics.forth.gr/lodsyndesis/rest-api}}.


\textbf{Rule B. Search for the same Subject-Predicate (SP) or Subject-Object (SO).} 
We search for finding triples that have either the same Subject-Predicate (SP) or the Same Subject-Object (SO) Pair, we collect all of them and we use them as candidates, i.e., see lines 4-5 in Alg. \ref{alg:valid}. 

{\em How the Rule is Executed?} For collecting the candidates of Rule B, for the case of DBpedia we can send two SPARQL SELECT queries, i.e., one with fixed subject-predicate and one with fixed subject-object. On the contrary, for the case of LODsyndesis, we can send two requests to its Fact Checking service.

\textbf{Rule C. Retrieve All Triples of $e$ as Candidates.}  If the previous two rules fail, we use all the triples of the subject $e$ of $t^{gpt}$ (where $e$ occurs either as a subject or as an object), i.e.,  $cand(t^{gpt})=T_{KG}(e)$ (see lines 6-7 of Alg. \ref{alg:valid}). 

{\em How the Rule is Executed?} For collecting the candidates, we either send a SELECT SPARQL query for retrieving all the facts for the entity $e$, or a  REST request to the ``allFacts" service of LODsyndesis REST API.

\subsubsection{Step 3.2 Creating the Embeddings.}
For the next steps (lines 8-13 of Alg. \ref{alg:valid}), we create a vector for both the $t^{gpt}$ and each $t^{gpt} \in cand(t^{gpt})$.

\textbf{Short Sentence Embeddings.} Since each triple is a short statement, and given the detected problems from \CG (see Fig. \ref{fig:context}), e.g., different predicates, literals instead of URIs, wrong and invalid URIs, including properties, etc.,  we decided to use sentence similarity models. Indeed, ``Sentence Similarity is the task of determining how similar two texts are, by converting input texts into vectors (embeddings) that capture semantic information and calculate how close (similar) they are between them"\footnote{\url{https://huggingface.co/tasks/sentence-similarity}}. 
Here, we use the most popular Sentence Similarity model from huggingface, called ``sentence-transformers/all-MiniLM-L6-v2" \cite{hugging2023}, which ``maps sentences and paragraphs to a 384 dimensional dense".
The objective is to measure how similar each candidate triple is with the fact $t^{gpt}$ in terms of their semantic meaning. However, such models need to process natural text and not triples, thereby, some URI conversions are needed.

\textbf{Conversions for URIs.} 
For the URIs of $t^{gpt}$ and of each   $t^{\prime} \in cand(t^{gpt})$, we perform some  conversions, i.e., we remove the prefixes and the special characters from the URIs such as the underscore ``\_". Moreover, in several cases, capital letters are used for distinguishing two or more words, e.g., dbo:associatedBand, and in such cases we split the words when a capital letter is identified, e.g., the previous will be converted to ``associated Band". On the contrary, for the URIs whose suffix is not human-readable (e.g., Wikidata), we replace the URIs with their labels. As an example,  the triple ``dbr:El\_Greco, dbo:artist dbr:View\_of\_Toledo", will be converted to ``El Greco artist View of Toledo", however, we also keep the initial triple and its provenance (i.e., for returning it as the final output).

\begin{table*}[ht]
\begin{tabular}{|p{1.1cm}|p{3.5cm}|r|>{\raggedleft\arraybackslash}p{2.1cm}|>{\raggedleft\arraybackslash}p{2.3cm}|>{\raggedleft\arraybackslash}p{2.5cm}|>{\raggedleft\arraybackslash}p{2.5cm}|}\hline
Part&Description of Entities that we asked \CG&Facts&Correct Facts (percentage)&Erroneous Facts (percentage)&Unique URIs (dereferencable)&Unique Properties   (dereferencable) \\\hline
Greek Persons&From the Great Greeks List&1000&812 (81.2\%)
&188 (18.8\%)&525 (472)&109 (85) \\\hline
Greek Places&Heritage Sites, Cities, Islands, Lakes, Mountains &500&319  (63.8\%) 
&181  (36.2\%)  &219 (177)& 85 (56)\\\hline
\new{Greek Events}&\new{Battles, Sports, Earthquakes, Elections}&500&330  (66.0\%) 
&170  (34.0\%)  &237 (204)& 140 (49)\\\hline
Total&-&2000
&1461 (73.05\%)&539 (26.95\%)&981 (853)&334 (190) \\\hline
\end{tabular}
\caption{The Evaluation Benchmark and Statistics}
\label{tbl:benchmark}
\end{table*}

\textbf{Creation of Vectors for the Converted Triples. } 
\new{Concerning the creation of embeddings, we give as input the converted KG triples to the "sentence-transformers/all-MiniLML6-v2” model; first}  we create the vector for the converted \CG fact $t^{gpt}_c$, i.e., we create $\vec{t^{gpt}_{c}}$ (lines 8-9). Afterwards, for each $t^{\prime} \in cand(t^{gpt})$, we create a vector for its converted version, i.e., $\vec{t^{\prime}_{c}}$ (lines 10-12). \new{At the end, each converted triple has been encoded to a vector of 384 dimensions. The motivation is to use these vectors to find the top-K triples with same/similar meaning to each \CG fact (according to their vectors’ cosine similarity), even if \CG used different URIs/Literals compared to the KG triples.} 

\subsubsection{Step 3.3 Computation of their Cosine Similarity with triple $t^{gpt}$.} For each $t^{\prime}$ we compute the cosine similarity with the triple $t^{gpt}$ (line 13). We selected to use cosine similarity, since it is has been successfully (and widely) used for RDF similarity-based approaches based on embeddings (e.g., see \cite{chatzakis2021rdfsim,ristoski2019rdf2vec}).
The cosine similarity ranges from -1 (i.e., the two vectors are exactly opposite) to 1 (i.e., the two vectors
are exactly the same, thereby the triples are equivalent).

\subsection{Step 4. Return the top $K$ similar triples (and provenance).}
The last step is to  sort the candidate triples according to their cosine similarity, for returning the top-$k$ results (i.e., see lines 14-15). Apart from the best triple(s), we return the cosine similarity score and the provenance of each triple. Concerning the value of $K$, some suggested values are $1,3,5$. Certainly, in some cases we have only one candidate, e.g., always in cases when Rule A is followed (e.g., see the example of Fig. \ref{fig:runningEx}), and in many cases either one or a few candidates, i.e., when Rule B is executed (e.g., see the example of Fig. \ref{fig:runningEx}).
On the contrary, for the rule C, we always have a larger number of candidates, since we compute the similarity between $t^{gpt}$ and all the  triples of the entity, e.g., see the example of Fig. \ref{fig:runningEx}. In that case, we successfully verified the \CG fact, although the KG used a different predicate and object URI for expressing the same real fact. 
\new{Finally, we should mention that the proposed pipeline spots the more similar facts of the KGs and provides the provenance of these facts, offering to the user a kind of support for the returned facts. This functioning mode does not presuppose having KGs with no conflicting values.}

\subsection{Time and Space Complexity of Alg. \ref{alg:valid}.} In the worst case, we need to  iterate over all the triples $T_{KG}(e)$ for the subject $e$ of the fact $t^{gpt}$, i.e., for creating the vector of each candidate and to compute its cosine similarity. Then we need to sort them according to their similarity score thereby the time complexity is $O(n*log(n))$, where $n$ refers to $T_{KG}(e)$ in our case. On the contrary, we load in memory the sentence similarity model, and we keep in memory all the candidate triples and their cosine similarity score, thereby the space complexity  is $O(n+m)$ ($n$ refers to $T_{KG}(e)$, and $m$ to the size of the sentence similarity model).



\begin{table*}[ht]
    \begin{minipage}{.33\linewidth}

      \centering
     \begin{tabular}{|p{0.17cm}|p{2.0cm}|>{\raggedleft\arraybackslash}p{0.6cm}|>{\raggedleft\arraybackslash}p{0.6cm}|>{\raggedleft\arraybackslash}p{0.6cm}|}\hline
ID&Predicate of Fact from \CG&Total Facts&Corr. Facts&Erron. Facts \\\hline
1&dbo:occupation&102&97&5 \\\hline
2&dbo:birthDate&90&60&30 \\\hline
3&rdf:type&85&85&0 \\\hline
4&dbo:deathDate&83&63&20 \\\hline
5&dbo:birthPlace&78&61&17 \\\hline
6&dbo:nationality&74&74&0 \\\hline
7&dbo:deathPlace&69&63&6 \\\hline
8&dbo:influenced&33&19&14 \\\hline
9&dbo:notableWork&21&17&4 \\\hline
10&dbo:field&20&17&3 \\\hline
\end{tabular}
      \caption{Top-10 Most Used Predicates for Greek Persons}
      \label{tbl:popularPersons}
    \end{minipage}%
    \hspace{2mm}
    \begin{minipage}{.31\linewidth}
      \centering
          \begin{tabular}{|p{0.17cm}|p{1.8cm}|>{\raggedleft\arraybackslash}p{0.6cm}|>{\raggedleft\arraybackslash}p{0.6cm}|>{\raggedleft\arraybackslash}p{0.6cm}|}\hline
ID&Predicate of Fact from \CG&Total Facts&Corr. Facts&Erron. Facts \\\hline
1&rdf:type&56&56&0 \\\hline
2&dbo:country&51&51&0 \\\hline
3&dbo:elevation&36&11&25 \\\hline
4&dbo:location&21&20&1 \\\hline
5&dbo:population&19&4&15 \\\hline
6&dbo:areaTotal&19&0&19 \\\hline
7&dbo:timeZone&17&17&0 \\\hline
8&dbo:settlement&17&14&3 \\\hline
9&dbo:locatedIn&15&15&0 \\\hline
10&dbo:length&14&1&13 \\\hline
\end{tabular}
  \caption{Top-10 Most Used Predicates for Greek Places}
         \label{tbl:popularPlaces}
    \end{minipage} 
   \hspace{2mm}
    \begin{minipage}{.32\linewidth}
      \centering
          \begin{tabular}{|p{0.17cm}|p{1.9cm}|>{\raggedleft\arraybackslash}p{0.6cm}|>{\raggedleft\arraybackslash}p{0.58cm}|>{\raggedleft\arraybackslash}p{0.58cm}|}\hline
ID&Predicate of Fact from \CG&Total Facts&Corr. Facts&Erron Facts \\\hline
1&dbo:date&52&29&23 \\\hline
2&dbo:commander&27&3&24 \\\hline
3&dbo:result&20&14&6 \\\hline
4&rdf:type&19&19&0 \\\hline
5&dbo:country&18&18&0 \\\hline
6&dbo:magnitude&15&7&8 \\\hline
7&dbo:casualties&14&6&8 \\\hline
8&dbo:depth&13&2&11 \\\hline
9&dbo:place&11&10&1 \\\hline
10&dbo:combatant&10&9&1 \\\hline
\end{tabular}
  \caption{Top-10 Most Used Predicates for Events related to Greece}
         \label{tbl:populaEvents}
    \end{minipage} 
\vspace{-2mm}
\end{table*}

\begin{table*}[!ht]
\begin{tabular}{|r|p{5cm}|p{5cm}|p{2cm}|p{2.3cm}|}\hline
\textbf{ID}&\textbf{\CG fact} & \textbf{Top Fact (LODsyndesis)} &\textbf{Annotation}&\textbf{Rule Used}\\\hline
1&dbr:Aristophanes dbo:genre dbr:Comedy&dbr:Aristophanes wkd:P136 dbr:Comedy&C1. Correct \& Validated&A (equivalent triple) \\\hline
2&dbr:Georgios\_Papanikolaou dbo:knownFor "Pap test"&dbr:Georgios\_Papanikolaou dbo:knownFor dbr:Pap\_Smear&C1. Correct \& Validated&B (same  predicate) \\\hline
3&dbo:Aristophanes dbo:notableWorks dbr:Lysistrata & dbo:Aristophanes dbo:author dbr:Lysistrata &C1. Correct \& Validated&B (same  object)\\\hline
4&dbr:Pericles dbo:office dbr:Strategos
&dbr:Pericles yago:rank "Strategos"&C1. Correct \& Validated&C (most similar triple) \\\hline \hline
5&dbr:Papadiamantis dbo:notableWork dbr:The\_Murderess&dbr:Papadiamantis rdf:type dbo:Writer &C2. Correct w/o Validation&C (most similar triples)\\\hline\hline
6&dbr:Georgios\_Papanikolaou dbo:birthDate "1886-05-13"& dbr:Georgios\_Papanikolaou dbo:birthDate "1883-05-13"&C3. Erroneous \& Validated&B (same predicate)\\\hline
7&dbr:Mikis\_Theodorakis dbo:activeYears "1949"& dbr:Mikis\_Theodorakis dbo:yearsActive "1943"&C3. Erroneous \& Validated&C (most similar triples)\\\hline \hline
8&dbr:Charilaos\_Florakis dbo:child dbr:Artemis\_Floraki&dbr:Charilaos\_Florakis yago:familyName ``florakis" &C4. Erroneous w/o Validation&C (most similar triples)\\\hline 
\end{tabular}
\caption{Real Examples of Labels (Annotations) from the Evaluation Benchmark}
\label{tbl:evalExamples}
\end{table*}

\section{Benchmark for \CG Facts \& RDF Knowledge Graphs} 
\label{sec:benchmark}
\new{ChatGPT was recently released, and to the best of our knowledge there is not any available collection of facts produced from ChatGPT. For this reason, we decided to create a benchmark  and make it publicly available, enabling anyone to use it for obtaining comparative results.  In principle, one could use an external dataset too, however that would not guarantee that ChatGPT has information about such facts, and thus it would not be suitable for validating the facts returned by ChatGPT. Concerning the benchmark, } we collected a list of famous Greek Persons, Places and \new{Events}\footnote{We selected facts that are related to Greece for facilitating the manual evaluation, since both authors are from Greece}. 
For the Persons, we used the list of the 100 Greatest Greeks of all time \cite{great}, including persons from different eras (Ancient and Modern era) and disciplines (leaders, artists, etc.). For the Places, we used a list of famous heritage sites and islands, and the largest cities, lakes and mountains. \new{Finally, for the Events, we used a list of battles, sports events, earthquakes and elections.}

\subsection{Collections of Facts from \CG and Manual Annotation.}
For each entity, we send the following requests to \CG (using ``gpt-3.5-turbo") 
as follows: ``Give me facts in RDF N-Triples format for entity $e$ using DBpedia format", where $e$ is replaced by the name of the person (like Aristotle), of the place (like Santorini) \new{or of the event (like 2004 Summer Olympics)}. The resulting benchmark contains \new{2,000} facts as it can be seen in Table \ref{tbl:benchmark}, which also shows some statistics about the benchmark. 

Then, we collected and manually labelled the facts of \CG as correct or erroneous (annotation conducted in Summer 2023). We manually annotated as:

$\bullet$ {\em Correct Facts}, the \CG facts that can be verified from online sources by searching on the web (e.g., Wikipedia, DBpedia, Wikidata, etc.). 

$\bullet$ {\em Erroneous Facts}  (including inaccurate ones, e.g., small differences in numbers), the \CG facts where we found the correct answer on the web (and it was a different one) or/and facts that cannot be verified from online sources, i.e., we did not find any evidence.

\subsection{Statistics over the Benchmark.} Table \ref{tbl:benchmark} introduces the collection by showing the number of facts, how many of them were correct or erroneous, the number of unique URIs and properties and whether the produced triples contain dereferencable URIs.   As we can see, the facts for the people were  more accurate compared to the Places, i.e., for the people approximately 1 out of 5 facts was erroneous, whereas for the Places and Events, approximately 1 out of 3. Concerning  URIs describing resources, 90\% for Persons, 80\% for Places \new{and 86\% of Events} are dereferencable, whereas for Properties, the corresponding percentages are 77\% for Persons, 66\% of Places \new{and only 35\% for Events}. For measuring whether a URI is dereferencable, we send an ASK query to DBpedia for checking if exists at least one triple including the desired URI (resource or property).

\subsection{Most Frequent Predicates, Correct and Erroneous \CG Facts} Concerning the most frequent predicates, for the Persons, i.e., see Table \ref{tbl:popularPersons}, they mainly describe general information such as the occupation of persons, their birth/death date and birth/death place and others. Concerning the erroneous facts, they were mainly facts related to numbers and dates, such as ``birth/death dates", but also some other ones like ``birth and death place" of a person. Concerning  other frequent erroneous cases, they include the predicates ``influenced", ``education" and "child". On the contrary, several correct facts were indicatively about the ``type", ``nationality", `notable works" and ``fields" of a person. 

Regarding the Places  (see Table \ref{tbl:popularPlaces}), the most erroneous \CG facts contain numbers, e.g., ``elevation", ``population", ``areaTotal", ``length", ``width", ``depth", whereas it also failed in many cases to find the ``major" of specific cities. On the contrary, \CG responses were more accurate for properties like the ``type", ``location", ``country" and ``timeZone". \new{Concerning the different types of places, the most erroneous facts were for Lakes (55\% erroneous), and the most correct ones for Cities (79\% were correct)}. 

\new{Concerning the Events (Table \ref{tbl:populaEvents}), there were several erroneous \CG facts for the ``commanders" of battles and again for facts including dates and numbers, e.g., ``date", ``magnitude", ``casualties", ``depth", etc. On the contrary, the information was more accurate for characteristics like ``place", ``country", and ``type". As regards the different types of Events, the most erroneous facts were about ``battles" (43\% were erroneous) and the most correct ones for the events about elections (73\% were correct).}

Finally, concerning other inaccurate general cases, they include ``owl:sameAs" relationships to RDF KGs that use identifiers for their URIs (e.g., Wikidata), and properties with identifiers as values, e.g.,  ``wikiPageRevisionID".

\begin{table*}[t]
\centering
\begin{tabular}{|l|>{\raggedleft\arraybackslash}p{2.8cm}|>{\raggedleft\arraybackslash}p{2.8cm}|>{\raggedleft\arraybackslash}p{3cm}|>{\raggedleft\arraybackslash}p{3.1cm}|}\hline
Collection and KG&\% of Correct \CG facts that Validated from KG (C1)&\% of Correct \CG facts w/o Validation from KG (C2)&\% of Erroneous \CG facts that Validated from KG (C3)&\% of Erroneous \CG facts w/o Validation from KG (C4)\\\hline \hline
All Facts (using DBpedia)&79.1\%&20.9\%&55.4\%&44.6\%\\\hline
All Facts (using LODsyndesis)&\textbf{85.3\%}&14.7\%&\textbf{58.0\%}&32.0\%\\\hline \hline
Persons (using DBpedia)&82.5\%&17.5\%&52.1\%&47.9\%\\\hline
Persons (using LODsyndesis)&\textbf{92.2\%}&7.8\%&\textbf{57.4\%}&42.6\%\\\hline\hline
Places (using DBpedia)&74.6\%&25.4\%&46.9\%&53.1\%\\\hline
Places (using LODsyndesis)&\textbf{77.1\%}&22.9\%&\textbf{49.1\%}&50.9\%\\\hline\hline
Events (using DBpedia)&75.1\%&24.9\%&\textbf{68.2\%}&31.8\%\\\hline
Events (using LODsyndesis)&\textbf{76.3\%}&23.7\%&\textbf{68.2\%}&31.8\%\\\hline
\end{tabular}
\caption{\new{Overall Percentages for the whole benchmark and specific parts of the benchmark}}
\label{tbl:overall}
\end{table*}

\subsection{Pages for Evaluation Benchmark, Source Code and Web Application.}
The source code (python), the evaluation benchmark, including its initial and its human labelled version, statistics and the experimental results (which are presented in \S \ref{sec:eval}) are available in a github page\footnote{\url{https://github.com/mountanton/GPToLODS_FactChecking}}. Moreover, the web application, where one can also validate any \CG fact (including any of the facts of the presented benchmark), is also available online\footnote{\url{https://demos.isl.ics.forth.gr/GPToLODS/}}.

\section{Experimental Evaluation}
\label{sec:eval}
Here, we use the evaluation benchmark of \S  \ref{sec:benchmark}, for evaluating mainly the effectiveness of the proposed approach; \neo{we also evaluate the efficiency of the approach}. Our target is also to compare the performance of using only a single KG, i.e., DBpedia versus more KGs, i.e., LODsyndesis (where DBpedia is also included). For performing the experiments we used a {\em Backend Python 3 Google Compute Engine from Google Colab} with 12.7 GB RAM and 107.7  GB disk space.


\subsection{The Process, Manual Labelling \& Metrics}
For each fact of the benchmark, we find the $K=3$ most similar triples. We decided to use $K=3$, since in many cases the top-1 result can be correct but not the desired one, especially for multi-valued predicates. For instance, in an erroneous fact saying that the birth place of a person is the city of ``Thessaloniki", the algorithm returned as the top similar fact from DBpedia that his birth place is ``Greece" and as a second similar that is the city of ``Larissa". Although both triples are correct, the second is the desired one for the mentioned case (i.e., city of Larissa belongs to Greece). However, remind that when Rule A is used, there is only one result, whereas when rule B is used, the algorithm (when $K=3$) can return either 1, 2 or 3 results. 
\new{For providing accurate results, we manually annotated each pair of a \CG triple and the returned answer(s) of the used KG. We cross-checked the annotations and we decided to use only non-ambiguous facts; in rare cases where the authors failed to agree on ambiguous facts, the mentioned facts (of the benchmark) were replaced by new non-ambiguous facts.} The 4 categories are listed below:

$\bullet$ C1.{\em Correct \& Validated}: Correct in \CG and validated from the KG, even by using a different predicate and object, i.e., see the IDs 1-4 in Table \ref{tbl:evalExamples}. For example, for ID 1, the triples are equivalent since both ``dbo:genre" and ``wkd:P136" refer to the characteristic ``genre".

$\bullet$ C2.{\em Correct w/o Validation}: Correct in \CG, but not validated from the KG (maybe the KG did not contain the fact or Alg. \ref{alg:valid} failed to find it), e.g., in ID 5 of Table \ref{tbl:evalExamples} the fact is correct, however, the KG does not include it.

$\bullet$  C3. {\em Erroneous \& Validated}: Erroneous in \CG, but the KG provided the correct answer for the same real fact, e.g., see the IDs 6-7 in Table \ref{tbl:evalExamples}.

$\bullet$ C4.{\em Erroneous w/o Validation}: Erroneous in \CG and the KG failed to provide the correct answer (maybe the KG did not contain the fact  or included an erroneous value for the fact, and cases where the algorithm failed to find it), e.g., suppose that for some persons can produce facts that they had children (although they did not). For instance, see the ID 8 in Table \ref{tbl:evalExamples}, where the fact is incorrect, however, LODsyndesis does not include a relevant fact, since the greek politician ``Charilaos Florakis" did not have children.
 
\textbf{Metrics.} We count the number of facts of each category, by using i) only DBpedia and ii) LODsyndesis. The goal is to evaluate the {\em RQ1}, i.e., we expect to observe high numbers for  C1 and C3 (i.e., the KG(s) provided the correct answer) and low numbers for C2 and C4 (i.e., the KG(s) failed to find the correct answer),  and the {\em RQ2}, i.e., we expect better results by using multiple RDF KGs.

\textbf{Note.} We should note that for (some very few) facts,  where in the top-$3$ triples we found contradicting values from many KGs (e.g., for the population of a place), we labelled them as not validated (categories C2 and C4).

\new{\subsection{Overall Evaluation}
Table \ref{tbl:overall} shows the percentages by using either only DBpedia or LODsyndesis, for all the facts of the benchmark, and for each of the parts of the benchmark (Persons, Places, Events). Concerning the whole benchmark, by using LODsyndesis, we managed to validate 85.3\% of the correct \CG facts (+6.2\% compared to DBpedia) and to find the correct answer for the 58.0\% of erroneous \CG facts (+2.6\% compared to DBpedia). 

\begin{table*}[t]
\centering
\begin{tabular}{|l|>{\raggedleft\arraybackslash}p{2cm}|>{\raggedleft\arraybackslash}p{2cm}|>{\raggedleft\arraybackslash}p{2cm}|>{\raggedleft\arraybackslash}p{2cm}||>{\raggedleft\arraybackslash}p{0.8cm}|}\hline
Rule&C1. Correct and Validated&C2. Correct w/o Validation&C3. Erroneous and Validated&C4. Erroneous w/o Validation&Total\\\hline
A. Same/Equivalent Triple&571&0&0&0&571\\\hline
B. Same Subject Predicate&217&37&238&35&524\\\hline
B. Same Subject Object&212&29&4&12&257\\\hline
C. Most Similar Triples&249&149&71&179&648\\\hline\hline
Total&1246&215&313&226&2000\\\hline
\end{tabular}
\caption{Analysis of Results (based on Rules) for the whole benchmark Using LODsyndesis}
\label{tbl:analysis}
\vspace{-2mm}
\end{table*}

\begin{table*}[t]
\centering
\begin{tabular}{|l|>{\raggedleft\arraybackslash}p{2cm}|>{\raggedleft\arraybackslash}p{2cm}|>{\raggedleft\arraybackslash}p{2cm}|>{\raggedleft\arraybackslash}p{2cm}||>{\raggedleft\arraybackslash}p{0.8cm}|}\hline
Rule&C1. Correct and Validated&C2. Correct w/o Validation&C3. Erroneous and Validated&C4. Erroneous w/o Validation&Total\\\hline
A. Same/Equivalent Triple&441&0&0&0&441\\\hline
B. Same Subject Predicate&202&48&232&19&501\\\hline
B. Same Subject Object&178&37&2&11&228\\\hline
C. Most Similar Triples&335&220&65&210&830\\\hline\hline
Total&1156&305&299&240&2000\\\hline
\end{tabular}
\caption{Analysis of Results (based on Rules) for the whole benchmark Using DBpedia}
\label{tbl:dbpediaAnalysis}
\vspace{-2mm}
\end{table*}

\textbf{Analysis of Rules and Labelled Results .} In Table \ref{tbl:analysis} we analyze the results by using LODsyndesis for the three rules. When rule A was executed all the facts were correct, whereas for rule B, in most of the cases, when we found the same SO, the \CG facts were correct. On the contrary, when we found the same SP, in many cases the fact was erroneous. Indeed, from the 313 erroneous facts that we found the correct answer, 238 of them were verified by using the mentioned rule. Finally, concerning Rule C, in several cases we found the correct answer (categories C1 and C3), however, it was also the most frequent rule when we failed to find the correct answer (i.e., approximately in 50\% of cases when that rule executed, we failed to find the correct answer in the KG). Regarding DBpedia, the results are presented in Table \ref{tbl:dbpediaAnalysis}. In that case, we can see that Rule A executed less times and Rule C more times compared to LODsyndesis KG.

\begin{figure*}[t]
    \centering
\includegraphics[width=110mm]{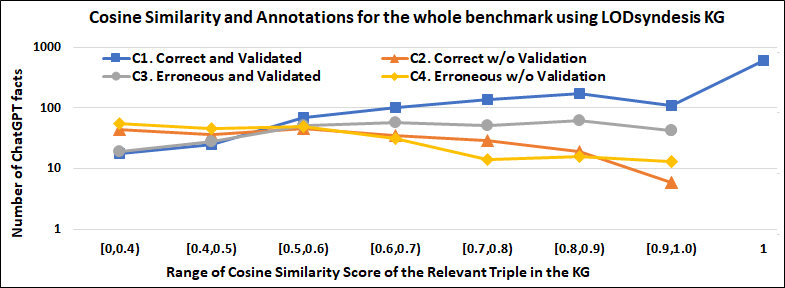}
\caption{Cosine Similarity and Annotations for the whole Benchmark using LODsyndesis}
    \label{fig:cosine}
\end{figure*}

\begin{figure*}[t]
    \centering
\includegraphics[width=110mm]{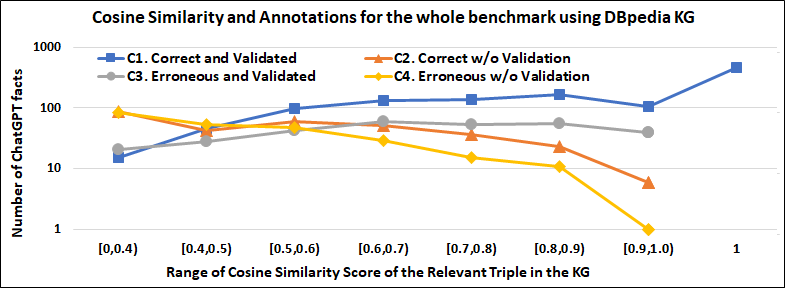}
\caption{Cosine Similarity and Annotations for the whole Benchmark using DBpedia}
    \label{fig:cosineDBpedia}
\end{figure*}

\textbf{Cosine Similarity and Annotations.}  Fig. \ref{fig:cosine} and Fig. \ref{fig:cosineDBpedia} depicts the number of \CG facts with respect to their category and the cosine similarity score of the relevant triple in the KG (returned by the algorithm), for LODsyndesis and DBpedia respectively. In both cases, we can see that for high cosine similarity scores, most facts were validated (see categories C1 and C3), whereas for low cosine similarity scores (between the \CG fact and the fact from the KG) less facts were verified. It is worth noting that in some cases, 
a fact in the KG was correct and in \CG erroneous (i.e. Category C3), however, their similarity was very high, e.g., for similar numbers, dates and places.

\textbf{Collection Parts.} As regards the specific collection parts, which are explained in the following subsections, we can observe in Table \ref{tbl:overall} that we managed to verify more correct facts for the Persons collection compared to the other ones, especially by using LODsyndesis KG. The main reason is that for the Persons collections, we found information from more KGs in LODsyndesis compared to Places and Events.
}

\begin{figure*}[t]
\begin{minipage}[b]{0.45\linewidth}
\centering
\includegraphics[width=0.9\textwidth]{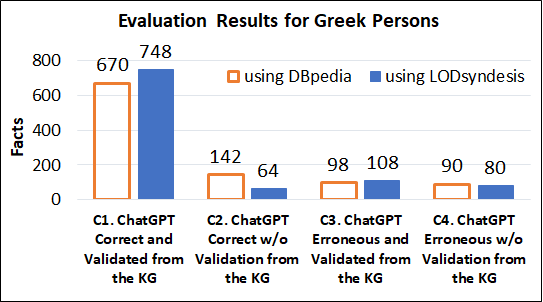}
\caption{Evaluation Results for {\em Persons} for all the categories of \CG facts (manually annotated)}
\label{fig:personsEval}
\end{minipage}
\hspace{0.4cm}
\begin{minipage}[b]{0.45\linewidth}
\centering
\includegraphics[width=0.9\textwidth]{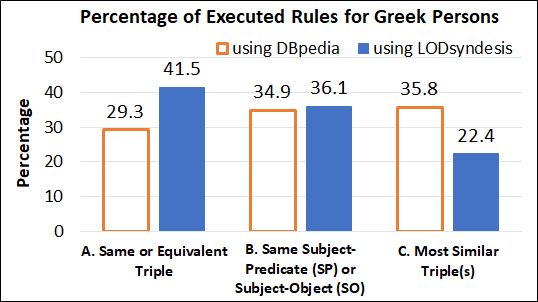}
\caption{Percentage of Rules Executed for {\em Persons} for both KGs}
\label{fig:personsRules}
\end{minipage}
\end{figure*}

\begin{figure*}[t]
\begin{minipage}[b]{0.45\linewidth}
\centering
\includegraphics[width=0.9\textwidth]{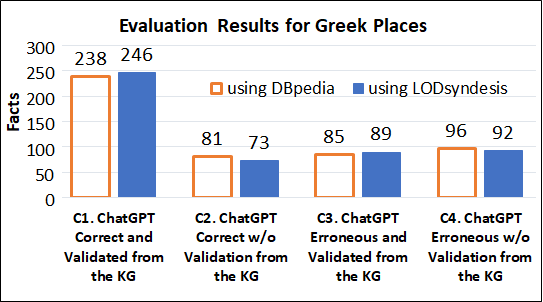}
\caption{Evaluation Results for  {\em Places}  for all the categories of \CG facts  (manually annotated)}
\label{fig:placesEval}
\end{minipage}
\hspace{0.5cm}
\begin{minipage}[b]{0.45\linewidth}
\centering
\includegraphics[width=0.9\textwidth]{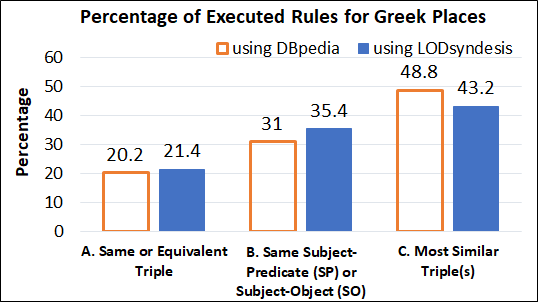}
\caption{Percentage of Rules Executed for {\em Places} for both KGs}
\label{fig:placesRules}
\end{minipage}
\end{figure*}

\new{

\begin{figure*}[t]
\begin{minipage}[b]{0.45\linewidth}
\centering
\includegraphics[width=0.9\textwidth]{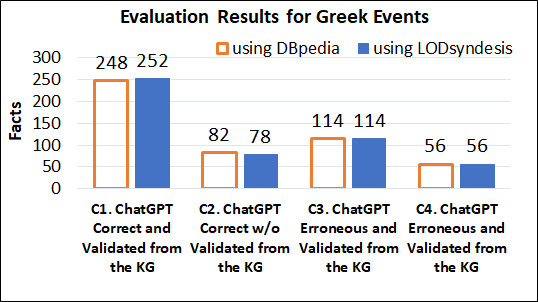}
\caption{Evaluation Results for  {\em Events} for all the categories of \CG facts  (manually annotated)}
\label{fig:eventsEval}
\end{minipage}
\hspace{0.5cm}
\begin{minipage}[b]{0.45\linewidth}
\centering
\includegraphics[width=0.9\textwidth]{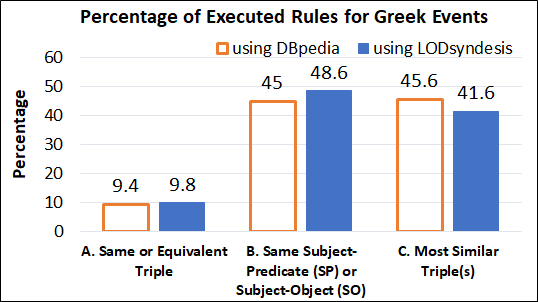}
\caption{Percentage of Rules Executed for {\em Events} for both KGs}
\label{fig:eventsRules}
\end{minipage}
\end{figure*}

\subsection{Validation of Facts for Persons} As we can see in Fig. \ref{fig:personsEval}  we managed to validate 748 correct facts (out of 812) by using LODsyndesis, i.e., 92.2\% of the correct \CG facts, whereas by using only DBpedia, we validated 670 correct facts, i.e., 82.5\%. As regards the erroneous \CG facts, we validated (i.e, we found the correct answer) 108 (out of 188) facts  by using LODsyndesis, i.e.,  57.4\%,  and 98 erroneous facts by using only DBpedia, i.e., 52.1\%.  By using LODsyndesis we managed to validate more facts, since some facts that are not described in DBpedia, were found in other popular KGs such as Wikidata \cite{vrandevcic2014wikidata}, YAGO \cite{rebele2016yago} and Freebase \cite{bollacker2008freebase}. This can be explained through Fig. \ref{fig:personsRules}, which shows the percentage of rules that were executed for each KG. In particular, by using LODsyndesis in 41.5\% of cases we found the same or equivalent triple (due to the precomputed closure), whereas the corresponding percentage for DBpedia was 29.3\%. Moreover, by using LODsyndesis in 36.1\% of cases we found at least the same SP or SO Pair and only in 22.4\% of cases the rule C executed. On the contrary, for DBpedia, in 34.9\% of cases we found the same SP or SO Pair, and in 35.8\% of cases the rule C executed.

\subsection{Validation of Facts for Places}
For this collection, the results are shown in Fig. \ref{fig:placesEval}. Concerning the correct facts we managed to verify 238 (out of 319) by using DBpedia, i.e., 74.6\%, and 246 of them by using LODsyndesis, i.e., 77.1\%. For the erroneous facts, we verified 85 (out of 181) through DBpedia (i.e., 46.9\%) and 89 of them using LODsyndesis (i.e., 49.1\%).  In that case, the difference is low, since less KGs (apart from DBpedia) that are included in LODsyndesis provided data for Places (compared to the Persons collection). Concerning the executed rules (see Fig \ref{fig:placesRules}) for this collection we found less equivalent triples and SP or SO Pairs, thereby, Rule C was executed more times compared to the Persons.

\subsection{Validation of  Facts for Events}
For this collection, the results are shown in Fig. \ref{fig:eventsEval}. Similarly to Places, the difference between using only DBpedia and LODsyndesis is quite small compared to the Persons, i.e., by using LODsyndesis we verified only 4 more correct \CG facts compared to using only DBpedia. On the contrary, it is worth noting that we managed to find the correct answer to a high percentage of erroneous \CG facts compared to Persons and Places. As regards the rules (see Fig \ref{fig:eventsRules}), we did not find several equivalent triples, and in most cases the Rules B and C were executed. The main reason is that compared to Persons and Places, only a small percentage of properties (returned by \CG) were dereferencable (i.e., see the last column of Table \ref{tbl:benchmark}). 
}

\begin{figure*}[t]
\begin{minipage}[b]{0.49\linewidth}
\centering
\includegraphics[width=0.87\textwidth]{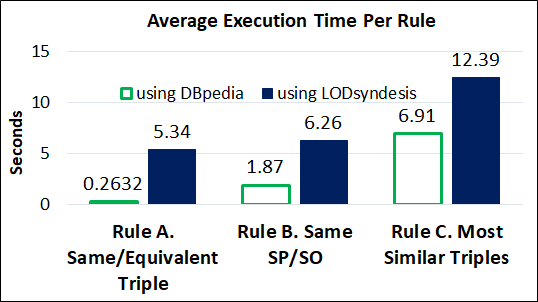}
\caption{Average Execution Time for each Rule and KG for the whole benchmark}
\label{fig:avgRules}
\end{minipage}
\hspace{0.1cm}
\begin{minipage}[b]{0.49\linewidth}
\centering
\includegraphics[width=0.87\textwidth]{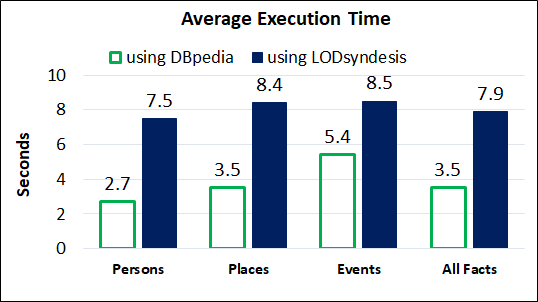}
\caption{Average Execution Time for each KG and each part of the evaluation benchmark}
\label{fig:avg}
\end{minipage}
\end{figure*}

\subsection{Discussion and Limitations}
Concerning the {\em RQs}, we verified most of the \CG facts  by using the proposed pipeline (i.e., {\em RQ1}), whereas by using multiple KGs we observed better results (i.e., {\em RQ2}), especially for the entities that belong to Persons. 
\new{For the correct  \CG facts, in most cases (i.e., 85.3\%) we validated the facts from the KG(s).
} As regards erroneous \CG facts that Alg. \ref{alg:valid} found the correct answer (i.e., 58.0 \%),  they mainly include facts about the birth/death date or place of a person, education of persons,  facts concerning the predicate influenced, the elevation of places, the major of cities, commanders of battles and others. 

On the contrary, for cases where we failed to find the correct answer in the KG(s), either for correct \CG facts (i.e., 14.7\% of cases) or for erroneous \CG facts (i.e., 37.4 \% of cases) the most common problems follow: a) the answer was included in large literals (e.g., abstracts), and the similarity score with such triples and the desired fact was low, b) predicates with very similar suffix, e.g., ``population",``populationTotal", ``populationAsOf", c) contradicting values among KGs (by using LODsyndesis), and d) incompleteness issues, i.e., cases where the KG(s) did not contain the fact, such as awards, missing birth date and place, intensity of earthquakes, the currency of a given city/island and notable work of persons.

For overcoming such limitations, a possible research direction could be to provide an extended version of Alg. \ref{alg:valid}, i) for supporting text extraction from large literals (e.g., through Question Answering techniques), ii) for resolving the conflicts between different KGs (e.g., for predicates and objects) and iii) for performing web search using a search engine (for tackling incompleteness issues).

\subsection{Efficiency Results}
Here, we provide efficiency measurements for evaluating the execution time  i)  for each rule and ii) for each different part of the evaluation benchmark, by using both KGs. 

First, Fig. \ref{fig:avgRules} shows the average execution time per rule for both KGs. As we can see, by using DBpedia it is quite fast to find the corresponding triples when Rules A and B are executed, since we just send some SPARQL queries. On the contrary, for Rule C, the execution time is much higher, since it also includes  the creation of the embeddings. On the contrary, by using LODsyndesis, more time is needed for all the rules, since i) its REST API is not so fast compared to the DBpedia SPARQL Endpoint and ii) it contains more triples for the same entities.

Second, Fig. \ref{fig:avg} shows the average execution time
for validating the \CG facts, by using a) DBpedia and b) LODsyndesis. Specifically, for the Persons the average execution time per fact was lower, since the rule C, which is the most time consuming one, was executed more times for the Places and Events collection.  For the whole benchmark, it was more than 2$\times$ faster to use only DBpedia, since it contains a smaller number of triples for each entity, compared to LODsyndesis (on average  3.5 sec versus 7.9 sec). Concerning the total time, by using DBpedia we needed 45 min for all the 1,000 facts  for Persons, 29 min for the 500 facts about Places and \new{45 min for the 500 facts about Events.} On the contrary, for LODsyndesis the corresponding time was 125 min for the Persons, 70 min for the Places, and \new{71 min for the Events.}

\section{Concluding Remarks}
\label{sec:conc}
Since \CG offers detailed responses without evidence, and  erroneous facts even for popular persons, events and places, in this paper we presented a novel pipeline that retrieves the response of \CG in RDF format (in N-triples format by using the DBpedia ontology) and tries to validate the \CG facts using one or more  RDF Knowledge Graphs (KGs).  For performing this task, we exploited DBpedia and LODsyndesis KG (which includes 400 RDF KGs), and we proposed an algorithm that is based on embeddings and short sentence similarity metrics. Moreover, we presented a web application that exploits the proposed pipeline and use cases, including  QA and Fact Validation, and Relation Extraction and Triples Generation. 

Afterwards, we created an evaluation benchmark including 2,000 facts derived from \CG for famous Greek persons, places and events, 
which were manually labelled; approximately 73\% of \CG facts were correct and 27\% of facts were erroneous. 
Concerning the results, by using LODsyndesis we verified 85.3\% of the correct \CG facts  (+6.2\% more compared to using only DBpedia). 
For the erroneous facts, we found the correct answer in 58.0\% of facts 
through LODsyndesis (+2.6\% versus DBpedia). \new{Regarding the efficiency, for validating each \CG fact, we needed on average 3.5 sec by using DBpedia and 7.9 sec by using LODsyndesis.}

\neo{Generally, we believe that such services for validation will be necessary for numerous applications that exploit Large Language Models (LLMs). Therefore,}
as a future work, we plan to i) further extend the benchmark with more entities, facts and domains,  \neo{from 
\CG and other LLMs}, ii) propose methods for automating the annotation/evaluation of \CG facts,  iii) extend the  algorithm for supporting answer extraction from large literals and web search,  iv) use/evaluate more sentence similarity models, and v) to evaluate the pipeline in responses of \new{other LLMs such as Google Bard \cite{bard}} or newer versions of \CG.


\balance 
\bibliographystyle{ACM-Reference-Format}
\bibliography{sample-base}


\begin{thebibliography}{48}


\ifx \showCODEN    \undefined \def \showCODEN     #1{\unskip}     \fi
\ifx \showDOI      \undefined \def \showDOI       #1{#1}\fi
\ifx \showISBNx    \undefined \def \showISBNx     #1{\unskip}     \fi
\ifx \showISBNxiii \undefined \def \showISBNxiii  #1{\unskip}     \fi
\ifx \showISSN     \undefined \def \showISSN      #1{\unskip}     \fi
\ifx \showLCCN     \undefined \def \showLCCN      #1{\unskip}     \fi
\ifx \shownote     \undefined \def \shownote      #1{#1}          \fi
\ifx \showarticletitle \undefined \def \showarticletitle #1{#1}   \fi
\ifx \showURL      \undefined \def \showURL       {\relax}        \fi
\providecommand\bibfield[2]{#2}
\providecommand\bibinfo[2]{#2}
\providecommand\natexlab[1]{#1}
\providecommand\showeprint[2][]{arXiv:#2}

\bibitem[Ammar and Celebi(2019)]%
        {ammar2019fact}
\bibfield{author}{\bibinfo{person}{Ammar Ammar} {and} \bibinfo{person}{Remzi
  Celebi}.} \bibinfo{year}{2019}\natexlab{}.
\newblock \showarticletitle{Fact Validation with Knowledge Graph Embeddings.}.
  In \bibinfo{booktitle}{\emph{ISWC (Satellites)}}. \bibinfo{pages}{125--128}.
\newblock


\bibitem[Beckett(2014)]%
        {beckett2014rdf}
\bibfield{author}{\bibinfo{person}{David Beckett}.}
  \bibinfo{year}{2014}\natexlab{}.
\newblock \showarticletitle{RDF 1.1 N-triples}.
\newblock \bibinfo{journal}{\emph{URL: https://www. w3. org/TR/n-triples}}
  (\bibinfo{year}{2014}).
\newblock


\bibitem[Biswas(2020)]%
        {biswas2020embedding}
\bibfield{author}{\bibinfo{person}{Russa Biswas}.}
  \bibinfo{year}{2020}\natexlab{}.
\newblock \showarticletitle{Embedding based link prediction for knowledge graph
  completion}. In \bibinfo{booktitle}{\emph{Proceedings of the 29th ACM
  International Conference on Information \& Knowledge Management}}.
  \bibinfo{pages}{3221--3224}.
\newblock


\bibitem[Bollacker et~al\mbox{.}(2008)]%
        {bollacker2008freebase}
\bibfield{author}{\bibinfo{person}{Kurt Bollacker}, \bibinfo{person}{Colin
  Evans}, \bibinfo{person}{Praveen Paritosh}, \bibinfo{person}{Tim Sturge},
  {and} \bibinfo{person}{Jamie Taylor}.} \bibinfo{year}{2008}\natexlab{}.
\newblock \showarticletitle{Freebase: a collaboratively created graph database
  for structuring human knowledge}. In \bibinfo{booktitle}{\emph{Proceedings of
  the 2008 ACM SIGMOD international conference on Management of data}}.
  \bibinfo{pages}{1247--1250}.
\newblock


\bibitem[Brown et~al\mbox{.}(2020)]%
        {brown2020language}
\bibfield{author}{\bibinfo{person}{Tom Brown}, \bibinfo{person}{Benjamin Mann},
  \bibinfo{person}{Nick Ryder}, \bibinfo{person}{Melanie Subbiah},
  \bibinfo{person}{Jared~D Kaplan}, \bibinfo{person}{Prafulla Dhariwal},
  \bibinfo{person}{Arvind Neelakantan}, \bibinfo{person}{Pranav Shyam},
  \bibinfo{person}{Girish Sastry}, \bibinfo{person}{Amanda Askell},
  {et~al\mbox{.}}} \bibinfo{year}{2020}\natexlab{}.
\newblock \showarticletitle{Language models are few-shot learners}.
\newblock \bibinfo{journal}{\emph{Advances in neural information processing
  systems}}  \bibinfo{volume}{33} (\bibinfo{year}{2020}),
  \bibinfo{pages}{1877--1901}.
\newblock


\bibitem[Chatzakis et~al\mbox{.}(2021)]%
        {chatzakis2021rdfsim}
\bibfield{author}{\bibinfo{person}{Manos Chatzakis}, \bibinfo{person}{Michalis
  Mountantonakis}, {and} \bibinfo{person}{Yannis Tzitzikas}.}
  \bibinfo{year}{2021}\natexlab{}.
\newblock \showarticletitle{RDFSIM: similarity-based browsing over dbpedia
  using embeddings}.
\newblock \bibinfo{journal}{\emph{Information}} \bibinfo{volume}{12},
  \bibinfo{number}{11} (\bibinfo{year}{2021}), \bibinfo{pages}{440}.
\newblock


\bibitem[Ciampaglia et~al\mbox{.}(2015)]%
        {ciampaglia2015computational}
\bibfield{author}{\bibinfo{person}{Giovanni~Luca Ciampaglia},
  \bibinfo{person}{Prashant Shiralkar}, \bibinfo{person}{Luis~M Rocha},
  \bibinfo{person}{Johan Bollen}, \bibinfo{person}{Filippo Menczer}, {and}
  \bibinfo{person}{Alessandro Flammini}.} \bibinfo{year}{2015}\natexlab{}.
\newblock \showarticletitle{Computational fact checking from knowledge
  networks}.
\newblock \bibinfo{journal}{\emph{PloS one}} \bibinfo{volume}{10},
  \bibinfo{number}{6} (\bibinfo{year}{2015}), \bibinfo{pages}{e0128193}.
\newblock


\bibitem[da~Silva et~al\mbox{.}(2021)]%
        {da2021using}
\bibfield{author}{\bibinfo{person}{Ana Alexandra~Morim da Silva},
  \bibinfo{person}{Michael R{\"o}der}, {and}
  \bibinfo{person}{Axel-Cyrille~Ngonga Ngomo}.}
  \bibinfo{year}{2021}\natexlab{}.
\newblock \showarticletitle{Using compositional embeddings for fact checking}.
  In \bibinfo{booktitle}{\emph{The Semantic Web--ISWC 2021: 20th International
  Semantic Web Conference, ISWC 2021, Virtual Event, October 24--28, 2021,
  Proceedings 20}}. Springer, \bibinfo{pages}{270--286}.
\newblock


\bibitem[Dong et~al\mbox{.}(2015)]%
        {dong2015knowledge}
\bibfield{author}{\bibinfo{person}{Xin~Luna Dong}, \bibinfo{person}{Evgeniy
  Gabrilovich}, \bibinfo{person}{Kevin Murphy}, \bibinfo{person}{Van Dang},
  \bibinfo{person}{Wilko Horn}, \bibinfo{person}{Camillo Lugaresi},
  \bibinfo{person}{Shaohua Sun}, {and} \bibinfo{person}{Wei Zhang}.}
  \bibinfo{year}{2015}\natexlab{}.
\newblock \showarticletitle{Knowledge-based trust: Estimating the
  trustworthiness of web sources}.
\newblock \bibinfo{journal}{\emph{arXiv preprint arXiv:1502.03519}}
  (\bibinfo{year}{2015}).
\newblock


\bibitem[Face(2023)]%
        {hugging2023}
\bibfield{author}{\bibinfo{person}{Hugging Face}.}
  \bibinfo{year}{2023}\natexlab{}.
\newblock \bibinfo{title}{sentence-transformers/all-MiniLM-L6-v2}.
\newblock
\newblock
\newblock
\shownote{\url{https://huggingface.co/sentence-transformers/all-MiniLM-L6-v2/}
  [Accessed: (September, 4, 2023)]}.


\bibitem[F{\"a}rber et~al\mbox{.}(2018)]%
        {farber2018linked}
\bibfield{author}{\bibinfo{person}{Michael F{\"a}rber},
  \bibinfo{person}{Frederic Bartscherer}, \bibinfo{person}{Carsten Menne},
  {and} \bibinfo{person}{Achim Rettinger}.} \bibinfo{year}{2018}\natexlab{}.
\newblock \showarticletitle{Linked data quality of dbpedia, freebase, opencyc,
  wikidata, and yago}.
\newblock \bibinfo{journal}{\emph{Semantic Web}} \bibinfo{volume}{9},
  \bibinfo{number}{1} (\bibinfo{year}{2018}), \bibinfo{pages}{77--129}.
\newblock


\bibitem[Geonames({[n.\,d.]})]%
        {geonames}
\bibfield{author}{\bibinfo{person}{Geonames}.}
  \bibinfo{year}{[n.\,d.]}\natexlab{}.
\newblock \bibinfo{title}{{GeoNames geographical database}}.
\newblock \bibinfo{howpublished}{\url{http://www.geonames.org/}}.
\newblock
\newblock
\shownote{Accessed: September, 4, 2023}.


\bibitem[Gerber et~al\mbox{.}(2015)]%
        {gerber2015defacto}
\bibfield{author}{\bibinfo{person}{Daniel Gerber}, \bibinfo{person}{Diego
  Esteves}, \bibinfo{person}{Jens Lehmann}, \bibinfo{person}{Lorenz
  B{\"u}hmann}, \bibinfo{person}{Ricardo Usbeck},
  \bibinfo{person}{Axel-Cyrille~Ngonga Ngomo}, {and} \bibinfo{person}{Ren{\'e}
  Speck}.} \bibinfo{year}{2015}\natexlab{}.
\newblock \showarticletitle{Defacto—temporal and multilingual deep fact
  validation}.
\newblock \bibinfo{journal}{\emph{Journal of Web Semantics}}
  \bibinfo{volume}{35} (\bibinfo{year}{2015}), \bibinfo{pages}{85--101}.
\newblock


\bibitem[Gonz{\'a}lez-Gallardo et~al\mbox{.}(2023)]%
        {gonzalez2023yes}
\bibfield{author}{\bibinfo{person}{Carlos-Emiliano Gonz{\'a}lez-Gallardo},
  \bibinfo{person}{Emanuela Boros}, \bibinfo{person}{Nancy Girdhar},
  \bibinfo{person}{Ahmed Hamdi}, \bibinfo{person}{Jose~G Moreno}, {and}
  \bibinfo{person}{Antoine Doucet}.} \bibinfo{year}{2023}\natexlab{}.
\newblock \showarticletitle{Yes but.. Can ChatGPT Identify Entities in
  Historical Documents?}
\newblock \bibinfo{journal}{\emph{arXiv preprint arXiv:2303.17322}}
  (\bibinfo{year}{2023}).
\newblock


\bibitem[{Google}(2023)]%
        {bard}
\bibfield{author}{\bibinfo{person}{{Google}}.} \bibinfo{year}{2023}\natexlab{}.
\newblock \bibinfo{booktitle}{\emph{Google Bard}}.
\newblock
\newblock
\shownote{Accessed: September 4, 2023}.


\bibitem[{Great Greeks}({[n.\,d.]})]%
        {great}
\bibfield{author}{\bibinfo{person}{{Great Greeks}}.}
  \bibinfo{year}{[n.\,d.]}\natexlab{}.
\newblock \bibinfo{title}{{Great Greeks}}.
\newblock
  \bibinfo{howpublished}{\url{https://en.wikipedia.org/wiki/Great_Greeks}}.
\newblock
\newblock
\shownote{Accessed: September, 4, 2023}.


\bibitem[Hoes et~al\mbox{.}(2023)]%
        {hoes2023using}
\bibfield{author}{\bibinfo{person}{Emma Hoes}, \bibinfo{person}{Sacha Altay},
  {and} \bibinfo{person}{Juan Bermeo}.} \bibinfo{year}{2023}\natexlab{}.
\newblock \showarticletitle{Using ChatGPT to Fight Misinformation: ChatGPT
  Nails 72\% of 12,000 Verified Claims}.
\newblock  (\bibinfo{year}{2023}).
\newblock


\bibitem[Hogan et~al\mbox{.}(2021)]%
        {hogan2021knowledge}
\bibfield{author}{\bibinfo{person}{Aidan Hogan}, \bibinfo{person}{Eva
  Blomqvist}, \bibinfo{person}{Michael Cochez}, \bibinfo{person}{Claudia
  d’Amato}, \bibinfo{person}{Gerard~de Melo}, \bibinfo{person}{Claudio
  Gutierrez}, \bibinfo{person}{Sabrina Kirrane}, \bibinfo{person}{Jos{\'e}
  Emilio~Labra Gayo}, \bibinfo{person}{Roberto Navigli},
  \bibinfo{person}{Sebastian Neumaier}, {et~al\mbox{.}}}
  \bibinfo{year}{2021}\natexlab{}.
\newblock \showarticletitle{Knowledge graphs}.
\newblock \bibinfo{journal}{\emph{ACM Computing Surveys (CSUR)}}
  \bibinfo{volume}{54}, \bibinfo{number}{4} (\bibinfo{year}{2021}),
  \bibinfo{pages}{1--37}.
\newblock


\bibitem[Huang et~al\mbox{.}(2022)]%
        {huang2022trustworthy}
\bibfield{author}{\bibinfo{person}{Jiacheng Huang}, \bibinfo{person}{Yao Zhao},
  \bibinfo{person}{Wei Hu}, \bibinfo{person}{Zhen Ning}, \bibinfo{person}{Qijin
  Chen}, \bibinfo{person}{Xiaoxia Qiu}, \bibinfo{person}{Chengfu Huo}, {and}
  \bibinfo{person}{Weijun Ren}.} \bibinfo{year}{2022}\natexlab{}.
\newblock \showarticletitle{Trustworthy knowledge graph completion based on
  multi-sourced noisy data}. In \bibinfo{booktitle}{\emph{Proceedings of the
  ACM Web Conference 2022}}. \bibinfo{pages}{956--965}.
\newblock


\bibitem[Huynh and Papotti(2018)]%
        {huynh2018towards}
\bibfield{author}{\bibinfo{person}{Viet-Phi Huynh} {and} \bibinfo{person}{Paolo
  Papotti}.} \bibinfo{year}{2018}\natexlab{}.
\newblock \showarticletitle{Towards a benchmark for fact checking with
  knowledge bases}. In \bibinfo{booktitle}{\emph{Companion Proceedings of the
  The Web Conference 2018}}. \bibinfo{pages}{1595--1598}.
\newblock


\bibitem[Huynh and Papotti(2019)]%
        {huynh2019buckle}
\bibfield{author}{\bibinfo{person}{Viet-Phi Huynh} {and} \bibinfo{person}{Paolo
  Papotti}.} \bibinfo{year}{2019}\natexlab{}.
\newblock \showarticletitle{Buckle: Evaluating fact checking algorithms built
  on knowledge bases}.
\newblock \bibinfo{journal}{\emph{Proceedings of the VLDB Endowment}}
  \bibinfo{volume}{12}, \bibinfo{number}{12} (\bibinfo{year}{2019}),
  \bibinfo{pages}{1798--1801}.
\newblock


\bibitem[Lehmann et~al\mbox{.}(2015)]%
        {lehmann2015dbpedia}
\bibfield{author}{\bibinfo{person}{Jens Lehmann}, \bibinfo{person}{Robert
  Isele}, \bibinfo{person}{Max Jakob}, \bibinfo{person}{Anja Jentzsch},
  \bibinfo{person}{Dimitris Kontokostas}, {et~al\mbox{.}}}
  \bibinfo{year}{2015}\natexlab{}.
\newblock \showarticletitle{Dbpedia--a large-scale, multilingual knowledge base
  extracted from wikipedia}.
\newblock \bibinfo{journal}{\emph{Semantic web}} \bibinfo{volume}{6},
  \bibinfo{number}{2} (\bibinfo{year}{2015}), \bibinfo{pages}{167--195}.
\newblock


\bibitem[Luo et~al\mbox{.}(2023)]%
        {luo2023chatrule}
\bibfield{author}{\bibinfo{person}{Linhao Luo}, \bibinfo{person}{Jiaxin Ju},
  \bibinfo{person}{Bo Xiong}, \bibinfo{person}{Yuan-Fang Li},
  \bibinfo{person}{Gholamreza Haffari}, {and} \bibinfo{person}{Shirui Pan}.}
  \bibinfo{year}{2023}\natexlab{}.
\newblock \showarticletitle{ChatRule: Mining Logical Rules with Large Language
  Models for Knowledge Graph Reasoning}.
\newblock \bibinfo{journal}{\emph{arXiv preprint arXiv:2309.01538}}
  (\bibinfo{year}{2023}).
\newblock


\bibitem[Maliaroudakis et~al\mbox{.}(2021)]%
        {maliaroudakis2021claimlinker}
\bibfield{author}{\bibinfo{person}{Evangelos Maliaroudakis},
  \bibinfo{person}{Katarina Boland}, \bibinfo{person}{Stefan Dietze},
  \bibinfo{person}{Konstantin Todorov}, \bibinfo{person}{Yannis Tzitzikas},
  {and} \bibinfo{person}{Pavlos Fafalios}.} \bibinfo{year}{2021}\natexlab{}.
\newblock \showarticletitle{ClaimLinker: Linking text to a knowledge graph of
  fact-checked claims}. In \bibinfo{booktitle}{\emph{Companion Proceedings of
  the Web Conference 2021}}. \bibinfo{pages}{669--672}.
\newblock


\bibitem[Mountantonakis(2021)]%
        {mountantonakis2021services}
\bibfield{author}{\bibinfo{person}{Michalis Mountantonakis}.}
  \bibinfo{year}{2021}\natexlab{}.
\newblock \bibinfo{booktitle}{\emph{Services for Connecting and Integrating Big
  Numbers of Linked Datasets}}. Vol.~\bibinfo{volume}{50}.
\newblock \bibinfo{publisher}{IOS Press}.
\newblock


\bibitem[Mountantonakis and Tzitzikas(2019)]%
        {mountantonakis2019knowledge}
\bibfield{author}{\bibinfo{person}{Michalis Mountantonakis} {and}
  \bibinfo{person}{Yannis Tzitzikas}.} \bibinfo{year}{2019}\natexlab{}.
\newblock \showarticletitle{Knowledge graph embeddings over hundreds of linked
  datasets}. In \bibinfo{booktitle}{\emph{Metadata and Semantic Research: 13th
  International Conference, MTSR 2019, Rome, Italy, October 28--31, 2019,
  Revised Selected Papers}}. Springer, \bibinfo{pages}{150--162}.
\newblock


\bibitem[Mountantonakis and Tzitzikas(2020)]%
        {mountantonakis2020content}
\bibfield{author}{\bibinfo{person}{Michalis Mountantonakis} {and}
  \bibinfo{person}{Yannis Tzitzikas}.} \bibinfo{year}{2020}\natexlab{}.
\newblock \showarticletitle{Content-based union and complement metrics for
  dataset search over RDF knowledge graphs}.
\newblock \bibinfo{journal}{\emph{ACM JDIQ}} \bibinfo{volume}{12},
  \bibinfo{number}{2} (\bibinfo{year}{2020}), \bibinfo{pages}{1--31}.
\newblock


\bibitem[Mountantonakis and Tzitzikas(2023a)]%
        {mountantonakis2023iswc}
\bibfield{author}{\bibinfo{person}{Michalis Mountantonakis} {and}
  \bibinfo{person}{Yannis Tzitzikas}.} \bibinfo{year}{2023}\natexlab{a}.
\newblock \showarticletitle{Real-Time Validation of ChatGPT facts using RDF
  Knowledge Graphs}.
\newblock \bibinfo{journal}{\emph{ISWC Demo Paper}} (\bibinfo{year}{2023}).
\newblock


\bibitem[Mountantonakis and Tzitzikas(2023b)]%
        {mountantonakis2023using}
\bibfield{author}{\bibinfo{person}{Michalis Mountantonakis} {and}
  \bibinfo{person}{Yannis Tzitzikas}.} \bibinfo{year}{2023}\natexlab{b}.
\newblock \showarticletitle{Using Multiple RDF Knowledge Graphs for Enriching
  ChatGPT Responses}.
\newblock \bibinfo{journal}{\emph{arXiv preprint arXiv:2304.05774}}
  (\bibinfo{year}{2023}).
\newblock


\bibitem[Omar et~al\mbox{.}(2023)]%
        {omar2023chatgpt}
\bibfield{author}{\bibinfo{person}{Reham Omar}, \bibinfo{person}{Omij
  Mangukiya}, \bibinfo{person}{Panos Kalnis}, {and} \bibinfo{person}{Essam
  Mansour}.} \bibinfo{year}{2023}\natexlab{}.
\newblock \showarticletitle{ChatGPT versus Traditional Question Answering for
  Knowledge Graphs: Current Status and Future Directions Towards Knowledge
  Graph Chatbots}.
\newblock \bibinfo{journal}{\emph{arXiv preprint arXiv:2302.06466}}
  (\bibinfo{year}{2023}).
\newblock


\bibitem[Ontotext(2023)]%
        {chatgpt2023rdf}
\bibfield{author}{\bibinfo{person}{Ontotext}.} \bibinfo{year}{2023}\natexlab{}.
\newblock \bibinfo{title}{Why should you combine ChatGPT with Knowledge
  Graphs?}
\newblock
\newblock
\newblock
\shownote{\url{https://www.ontotext.com/blog/why-should-you-combine-chatgpt-with-knowledge-graphs/}
  [accessed: (September, 4, 2023)]}.


\bibitem[{OpenAI}(2021)]%
        {OpenAI}
\bibfield{author}{\bibinfo{person}{{OpenAI}}.} \bibinfo{year}{2021}\natexlab{}.
\newblock \bibinfo{booktitle}{\emph{ChatGPT}}.
\newblock
\newblock
\shownote{Accessed: March 21, 2023}.


\bibitem[Pister and Atemezing(2019)]%
        {pister2019knowledge}
\bibfield{author}{\bibinfo{person}{Alexis Pister} {and}
  \bibinfo{person}{Ghislain~Auguste Atemezing}.}
  \bibinfo{year}{2019}\natexlab{}.
\newblock \showarticletitle{Knowledge Graph Embedding for Triples Fact
  Validation.}. In \bibinfo{booktitle}{\emph{ISWC (Satellites)}}.
  \bibinfo{pages}{21--24}.
\newblock


\bibitem[Portisch et~al\mbox{.}(2022)]%
        {portisch2022knowledge}
\bibfield{author}{\bibinfo{person}{Jan Portisch}, \bibinfo{person}{Nicolas
  Heist}, {and} \bibinfo{person}{Heiko Paulheim}.}
  \bibinfo{year}{2022}\natexlab{}.
\newblock \showarticletitle{Knowledge graph embedding for data mining vs.
  knowledge graph embedding for link prediction--two sides of the same coin?}
\newblock \bibinfo{journal}{\emph{Semantic Web}} \bibinfo{volume}{13},
  \bibinfo{number}{3} (\bibinfo{year}{2022}), \bibinfo{pages}{399--422}.
\newblock


\bibitem[Qudus et~al\mbox{.}(2022)]%
        {qudus2022hybridfc}
\bibfield{author}{\bibinfo{person}{Umair Qudus}, \bibinfo{person}{Michael
  R{\"o}der}, \bibinfo{person}{Muhammad Saleem}, {and}
  \bibinfo{person}{Axel-Cyrille Ngonga~Ngomo}.}
  \bibinfo{year}{2022}\natexlab{}.
\newblock \showarticletitle{HybridFC: A Hybrid Fact-Checking Approach for
  Knowledge Graphs}. In \bibinfo{booktitle}{\emph{The Semantic Web--ISWC 2022:
  21st International Semantic Web Conference, Virtual Event, October 23--27,
  2022, Proceedings}}. Springer, \bibinfo{pages}{462--480}.
\newblock


\bibitem[Rebele et~al\mbox{.}(2016)]%
        {rebele2016yago}
\bibfield{author}{\bibinfo{person}{Thomas Rebele}, \bibinfo{person}{Fabian
  Suchanek}, \bibinfo{person}{Johannes Hoffart}, \bibinfo{person}{Joanna
  Biega}, \bibinfo{person}{Erdal Kuzey}, {and} \bibinfo{person}{Gerhard
  Weikum}.} \bibinfo{year}{2016}\natexlab{}.
\newblock \showarticletitle{YAGO: A multilingual knowledge base from wikipedia,
  wordnet, and geonames}. In \bibinfo{booktitle}{\emph{The Semantic Web--ISWC
  2016: 15th International Semantic Web Conference, Kobe, Japan, October
  17--21, 2016, Proceedings, Part II 15}}. Springer, \bibinfo{pages}{177--185}.
\newblock


\bibitem[Ristoski et~al\mbox{.}(2019)]%
        {ristoski2019rdf2vec}
\bibfield{author}{\bibinfo{person}{Petar Ristoski}, \bibinfo{person}{Jessica
  Rosati}, \bibinfo{person}{Tommaso Di~Noia}, \bibinfo{person}{Renato
  De~Leone}, {and} \bibinfo{person}{Heiko Paulheim}.}
  \bibinfo{year}{2019}\natexlab{}.
\newblock \showarticletitle{RDF2Vec: RDF graph embeddings and their
  applications}.
\newblock \bibinfo{journal}{\emph{Semantic Web}} \bibinfo{volume}{10},
  \bibinfo{number}{4} (\bibinfo{year}{2019}), \bibinfo{pages}{721--752}.
\newblock


\bibitem[Shi et~al\mbox{.}(2023)]%
        {shi2023chatgraph}
\bibfield{author}{\bibinfo{person}{Yucheng Shi}, \bibinfo{person}{Hehuan Ma},
  \bibinfo{person}{Wenliang Zhong}, \bibinfo{person}{Gengchen Mai},
  \bibinfo{person}{Xiang Li}, \bibinfo{person}{Tianming Liu}, {and}
  \bibinfo{person}{Junzhou Huang}.} \bibinfo{year}{2023}\natexlab{}.
\newblock \showarticletitle{ChatGraph: Interpretable Text Classification by
  Converting ChatGPT Knowledge to Graphs}.
\newblock \bibinfo{journal}{\emph{arXiv preprint arXiv:2305.03513}}
  (\bibinfo{year}{2023}).
\newblock


\bibitem[Shiralkar et~al\mbox{.}(2017)]%
        {shiralkar2017finding}
\bibfield{author}{\bibinfo{person}{Prashant Shiralkar},
  \bibinfo{person}{Alessandro Flammini}, \bibinfo{person}{Filippo Menczer},
  {and} \bibinfo{person}{Giovanni~Luca Ciampaglia}.}
  \bibinfo{year}{2017}\natexlab{}.
\newblock \showarticletitle{Finding streams in knowledge graphs to support fact
  checking}. In \bibinfo{booktitle}{\emph{2017 IEEE International Conference on
  Data Mining (ICDM)}}. IEEE, \bibinfo{pages}{859--864}.
\newblock


\bibitem[Tan et~al\mbox{.}(2023)]%
        {tan2023evaluation}
\bibfield{author}{\bibinfo{person}{Yiming Tan}, \bibinfo{person}{Dehai Min},
  \bibinfo{person}{Yu Li}, \bibinfo{person}{Wenbo Li}, \bibinfo{person}{Nan
  Hu}, \bibinfo{person}{Yongrui Chen}, {and} \bibinfo{person}{Guilin Qi}.}
  \bibinfo{year}{2023}\natexlab{}.
\newblock \showarticletitle{Evaluation of ChatGPT as a question answering
  system for answering complex questions}.
\newblock \bibinfo{journal}{\emph{arXiv preprint arXiv:2303.07992}}
  (\bibinfo{year}{2023}).
\newblock


\bibitem[Trajanoska et~al\mbox{.}(2023)]%
        {trajanoska2023enhancing}
\bibfield{author}{\bibinfo{person}{Milena Trajanoska}, \bibinfo{person}{Riste
  Stojanov}, {and} \bibinfo{person}{Dimitar Trajanov}.}
  \bibinfo{year}{2023}\natexlab{}.
\newblock \showarticletitle{Enhancing Knowledge Graph Construction Using Large
  Language Models}.
\newblock \bibinfo{journal}{\emph{arXiv preprint arXiv:2305.04676}}
  (\bibinfo{year}{2023}).
\newblock


\bibitem[van Dis et~al\mbox{.}(2023)]%
        {van2023chatgpt}
\bibfield{author}{\bibinfo{person}{Eva~AM van Dis}, \bibinfo{person}{Johan
  Bollen}, \bibinfo{person}{Willem Zuidema}, \bibinfo{person}{Robert van
  Rooij}, {and} \bibinfo{person}{Claudi~L Bockting}.}
  \bibinfo{year}{2023}\natexlab{}.
\newblock \showarticletitle{ChatGPT: five priorities for research}.
\newblock \bibinfo{journal}{\emph{Nature}} \bibinfo{volume}{614},
  \bibinfo{number}{7947} (\bibinfo{year}{2023}), \bibinfo{pages}{224--226}.
\newblock


\bibitem[Vassiliou et~al\mbox{.}(2023)]%
        {summarygpt}
\bibfield{author}{\bibinfo{person}{Giannis Vassiliou},
  \bibinfo{person}{Nikolaos Papadakis}, {and} \bibinfo{person}{Haridimos
  Kondylakis}.} \bibinfo{year}{2023}\natexlab{}.
\newblock \showarticletitle{Summary{GPT}: Leveraging {ChatGPT} for Summarizing
  Knowledge Graphs}.
\newblock \bibinfo{journal}{\emph{ESCW}} (\bibinfo{year}{2023}).
\newblock


\bibitem[Vedula and Parthasarathy(2021)]%
        {vedula2021face}
\bibfield{author}{\bibinfo{person}{Nikhita Vedula} {and}
  \bibinfo{person}{Srinivasan Parthasarathy}.} \bibinfo{year}{2021}\natexlab{}.
\newblock \showarticletitle{Face-keg: Fact checking explained using knowledge
  graphs}. In \bibinfo{booktitle}{\emph{Proceedings of the 14th ACM
  International Conference on Web Search and Data Mining}}.
  \bibinfo{pages}{526--534}.
\newblock


\bibitem[Vrande{\v{c}}i{\'c} and Kr{\"o}tzsch(2014)]%
        {vrandevcic2014wikidata}
\bibfield{author}{\bibinfo{person}{Denny Vrande{\v{c}}i{\'c}} {and}
  \bibinfo{person}{Markus Kr{\"o}tzsch}.} \bibinfo{year}{2014}\natexlab{}.
\newblock \showarticletitle{Wikidata: a free collaborative knowledgebase}.
\newblock \bibinfo{journal}{\emph{Commun. ACM}} \bibinfo{volume}{57},
  \bibinfo{number}{10} (\bibinfo{year}{2014}), \bibinfo{pages}{78--85}.
\newblock


\bibitem[Yang et~al\mbox{.}(2023)]%
        {yang2023chatgpt}
\bibfield{author}{\bibinfo{person}{Linyao Yang}, \bibinfo{person}{Hongyang
  Chen}, \bibinfo{person}{Zhao Li}, \bibinfo{person}{Xiao Ding}, {and}
  \bibinfo{person}{Xindong Wu}.} \bibinfo{year}{2023}\natexlab{}.
\newblock \showarticletitle{ChatGPT is not Enough: Enhancing Large Language
  Models with Knowledge Graphs for Fact-aware Language Modeling}.
\newblock \bibinfo{journal}{\emph{arXiv preprint arXiv:2306.11489}}
  (\bibinfo{year}{2023}).
\newblock


\bibitem[Yao et~al\mbox{.}(2023)]%
        {yao2023exploring}
\bibfield{author}{\bibinfo{person}{Liang Yao}, \bibinfo{person}{Jiazhen Peng},
  \bibinfo{person}{Chengsheng Mao}, {and} \bibinfo{person}{Yuan Luo}.}
  \bibinfo{year}{2023}\natexlab{}.
\newblock \showarticletitle{Exploring Large Language Models for Knowledge Graph
  Completion}.
\newblock \bibinfo{journal}{\emph{arXiv preprint arXiv:2308.13916}}
  (\bibinfo{year}{2023}).
\newblock


\bibitem[Zeng et~al\mbox{.}(2021)]%
        {zeng2021automated}
\bibfield{author}{\bibinfo{person}{Xia Zeng}, \bibinfo{person}{Amani~S
  Abumansour}, {and} \bibinfo{person}{Arkaitz Zubiaga}.}
  \bibinfo{year}{2021}\natexlab{}.
\newblock \showarticletitle{Automated fact-checking: A survey}.
\newblock \bibinfo{journal}{\emph{Language and Linguistics Compass}}
  \bibinfo{volume}{15}, \bibinfo{number}{10} (\bibinfo{year}{2021}),
  \bibinfo{pages}{e12438}.
\newblock


\end{thebibliography}


\end{document}